\documentclass[aps,pr,onecolumn,superscriptaddress,preprintnumbers,showpacs,amsmath,amssymb]{revtex4}

\usepackage{graphicx} 
\usepackage{dcolumn}  
\usepackage{color}

\renewcommand{\arraystretch}{1.1}
\usepackage{epsfig}
\usepackage{epsbox}
\usepackage[T1]{fontenc}
\usepackage{wrapfloat}
\usepackage{subfigure}
\usepackage{here}
\usepackage{array}
\usepackage{amsmath, amssymb, hhline, multirow, hangcaption, subfigure}
\usepackage{graphics}
\usepackage{graphicx}
\usepackage{lscape}
\usepackage{fancyhdr}
\usepackage{longtable}
\usepackage{supertabular}
\usepackage{refmerge}

\def \vec #1{\mbox{{\boldmath $#1$}}}

\def \B {{\cal B}}

\def \GeV {{\rm GeV}}
\def \MeV {{\rm MeV}}

\def \eV {{\rm eV}}

\begin{document}


\title
{High-statistics study of ${\boldmath \eta \pi^0}$ production
in two-photon collisions}

\begin{abstract}
The differential cross section for the process $\gamma \gamma \to \eta \pi^0$
has been measured in the kinematic range
$0.84~\GeV < W < 4.0~\GeV$, $|\cos \theta^*|<0.8$,
where $W$ and $\theta^*$ are the energy and $\pi^0$ (or $\eta$) scattering 
angle, respectively, in the $\gamma\gamma$ center-of-mass system.
The results are based on a 223~fb$^{-1}$ data sample
collected with the Belle detector at the KEKB $e^+ e^-$ collider.
Clear peaks due to the $a_0(980)$ and $a_2(1320)$ are visible.
The differential cross sections are fitted in the energy region  
$0.9~\GeV < W <  1.46~\GeV$ to obtain the parameters of the $a_0(980)$.
Its mass, width and $\Gamma_{\gamma \gamma} \B (\eta \pi^0)$
are measured to be $982.3~^{+0.6}_{-0.7}~^{+3.1}_{-4.7}~\MeV/c^2$,
$75.6 \pm 1.6~^{+17.4}_{-10.0}~\MeV$ and
$128~^{+3}_{-2}~^{+502}_{-43}~\eV$, respectively.
The energy and angular dependences above 3.1~GeV are compared 
with those measured in the $\pi^0 \pi^0$ channel.
The integrated cross section over $|\cos \theta^*|<0.8$
has a $W^{-n}$ dependence 
with $n = 10.5 \pm 1.2 \pm 0.5$, which is slightly larger than that
for $\pi^0 \pi^0$.
The differential cross sections show a $\sin^{-4} \theta^*$ dependence
similar to $\gamma \gamma \to \pi^0 \pi^0$.
The measured cross section ratio, $\sigma(\eta \pi^0)/\sigma(\pi^0 \pi^0)
= 0.48 \pm 0.05 \pm 0.04$,
is consistent with a QCD-based prediction.
\end{abstract}

\normalsize
\affiliation{Budker Institute of Nuclear Physics, Novosibirsk}
\affiliation{Chiba University, Chiba}
\affiliation{University of Cincinnati, Cincinnati, Ohio 45221}
\affiliation{The Graduate University for Advanced Studies, Hayama}
\affiliation{Hanyang University, Seoul}
\affiliation{University of Hawaii, Honolulu, Hawaii 96822}
\affiliation{High Energy Accelerator Research Organization (KEK), Tsukuba}
\affiliation{Hiroshima Institute of Technology, Hiroshima}
\affiliation{University of Illinois at Urbana-Champaign, Urbana, Illinois 61801}
\affiliation{Institute of High Energy Physics, Chinese Academy of Sciences, Beijing}
\affiliation{Institute of High Energy Physics, Vienna}
\affiliation{Institute of High Energy Physics, Protvino}
\affiliation{Institute for Theoretical and Experimental Physics, Moscow}
\affiliation{J. Stefan Institute, Ljubljana}
\affiliation{Kanagawa University, Yokohama}
\affiliation{Korea University, Seoul}
\affiliation{Kyungpook National University, Taegu}
\affiliation{\'Ecole Polytechnique F\'ed\'erale de Lausanne (EPFL), Lausanne}
\affiliation{University of Maribor, Maribor}
\affiliation{Max-Planck-Institute for Physics, Munick}
\affiliation{University of Melbourne, School of Physics, Victoria 3010}
\affiliation{Nagoya University, Nagoya}
\affiliation{Nara Women's University, Nara}
\affiliation{National Central University, Chung-li}
\affiliation{National United University, Miao Li}
\affiliation{Department of Physics, National Taiwan University, Taipei}
\affiliation{H. Niewodniczanski Institute of Nuclear Physics, Krakow}
\affiliation{Nippon Dental University, Niigata}
\affiliation{Niigata University, Niigata}
\affiliation{University of Nova Gorica, Nova Gorica}
\affiliation{Novosibirsk State University, Novosibirsk}
\affiliation{Osaka City University, Osaka}
\affiliation{Panjab University, Chandigarh}
\affiliation{Saga University, Saga}
\affiliation{University of Science and Technology of China, Hefei}
\affiliation{Sungkyunkwan University, Suwon}
\affiliation{University of Sydney, Sydney, New South Wales}
\affiliation{Toho University, Funabashi}
\affiliation{Tohoku Gakuin University, Tagajo}
\affiliation{Tohoku University, Sendai}
\affiliation{Department of Physics, University of Tokyo, Tokyo}
\affiliation{Tokyo Metropolitan University, Tokyo}
\affiliation{Tokyo University of Agriculture and Technology, Tokyo}
\affiliation{IPNAS, Virginia Polytechnic Institute and State University, Blacksburg, Virginia 24061}
\affiliation{Yonsei University, Seoul}
  \author{S.~Uehara}\affiliation{High Energy Accelerator Research Organization (KEK), Tsukuba} 
  \author{Y.~Watanabe}\affiliation{Kanagawa University, Yokohama} 
 \author{H.~Nakazawa}\affiliation{National Central University, Chung-li} 
  \author{I.~Adachi}\affiliation{High Energy Accelerator Research Organization (KEK), Tsukuba} 
  \author{H.~Aihara}\affiliation{Department of Physics, University of Tokyo, Tokyo} 
  \author{K.~Arinstein}\affiliation{Budker Institute of Nuclear Physics, Novosibirsk}\affiliation{Novosibirsk State University, Novosibirsk} 
  \author{V.~Aulchenko}\affiliation{Budker Institute of Nuclear Physics, Novosibirsk}\affiliation{Novosibirsk State University, Novosibirsk} 
  \author{A.~M.~Bakich}\affiliation{University of Sydney, Sydney, New South Wales} 
  \author{K.~Belous}\affiliation{Institute of High Energy Physics, Protvino} 
  \author{A.~Bondar}\affiliation{Budker Institute of Nuclear Physics, Novosibirsk}\affiliation{Novosibirsk State University, Novosibirsk} 
% ...
\collaboration{The Belle Collaboration}


{\renewcommand{\thefootnote}{\fnsymbol{footnote}}

\setcounter{footnote}{0}
\setcounter{figure}{0}

\normalsize

\maketitle
\normalsize

\section{Introduction}
\label{sec-1}
  Measurements of exclusive hadronic final states in two-photon
collisions provide valuable information concerning physics of light and 
heavy-quark resonances, perturbative and non-perturbative QCD 
and hadron-production mechanisms.
So far, we have measured the production cross sections for 
charged-pion pairs~\cite{mori1,mori2,nkzw},
charged and neutral-kaon pairs~\cite{nkzw,kabe,wtchen},
and proton-antiproton pairs~\cite{kuo}.
We have also analyzed $D$-meson-pair production and observe a new
charmonium state identified as the $\chi_{c2}(2P)$~\cite{uehara}.
In addition, we have measured the $\pi^0 \pi^0$ final 
state~\cite{pi0pi0,pi0pi02}.
The statistics of these measurements is two to three orders of
magnitude higher than pre-$B$-factory measurements~\cite{past_exp}, 
opening a new era in studies of two-photon physics.

In this paper, we report measurements of the differential cross sections,
$d\sigma/d|\cos \theta^*|$,
for the process $\gamma \gamma \to \eta \pi^0$ in
a wide two-photon center-of-mass (c.m.) energy ($W$) range 
from 0.84~GeV to 4.0~GeV and
in the c.m. angular range, $|\cos \theta^*| \leq 0.8$.
We use only the
$\eta \to \gamma \gamma$ and $\pi^0 \to \gamma \gamma$ decay modes
in this analysis.
The decay mode $\eta \to \pi^+\pi^-\pi^0$ is not
used because of a much lower product of efficiency and
branching fraction.

Previously, it was reported that this reaction is dominated by resonance 
production~\cite{cbep}.
We can restrict  the $I^G J^{PC}$ quantum numbers
of the meson produced by two photons to be
$1^-$(even)$^{++}$, that is, those of $a_{J={\rm even}}$ mesons.
A long-standing puzzle in QCD
is the existence and structure of low mass scalar mesons.
In the $I=0$
sector, we recently observed a peak for the $f_0(980)$ in both
the $\gamma\gamma \to \pi^+\pi^-$ and $\gamma\gamma \to \pi^0\pi^0$ 
channels~\cite{mori1,pi0pi0}.
The two-photon width of the $f_0(980)$ is measured to be $200 - 300$~eV,
supporting its $q^2 \bar{q^2}$ nature~\cite{achasov}.
Our analysis also suggests the existence of another $f_0$ meson
in the 1.2-1.5~GeV region that couples with two photons~\cite{pi0pi0}.
In the $I=1$ sector, the $a_0(980)$ and $a_2(1320)$ have
been observed previously with a rather low statistical 
significance~\cite{cbep,jade}.
The parameters for the $a_0(980)$, in particular its two-photon width
are of great interest because of its connection to
the nature of low mass scalar mesons.
Moreover, other scalar or tensor $a_J$ mesons can be searched for
in the higher mass region above the prominent peak from
the $a_2(1320)$.

If there were an $I=1$ ``hidden-charm'' (that is, charmonium-like)
meson, it could be a very strong candidate for 
an exotic state, because charmonia
have $I=0$ and isospin is conserved in their hadronic decays.
However,
recently, some new particles that are not pure
$I=0$, such as the $X(3872)$~\cite{x3872} 
and $Z(4430)$~\cite{z4430}, have been reported.

At higher energies ($W > 2.4$~GeV), we can invoke a quark model.
In leading-order calculations, the ratio of the $\eta \pi^0$ cross section 
to that of $\pi^0 \pi^0$ is predicted within 
uncertainties due to the different form factors for the $\pi^0$ and $\eta$.
Analyses of energy and angular distributions of these cross sections
are essential to determine properties of the observed
resonances and to test the validity of QCD based models~\cite{bl,handbag}.

This paper is organized as follows.
In Sec.~\ref{sec-2}, the experimental apparatus used and the event 
selection are described.
Section \ref{sec-3} explains background subtraction and derivation of
the differential cross sections.
In Sec.~\ref{sec-4} the resonance parameters of the $a_0(980)$
are derived by parameterizing partial wave amplitudes with resonances and 
smooth non-resonant background amplitudes and fitting differential 
cross sections.
Section~\ref{sec-5} describes analyses at higher energy.
The topics included there are the angular dependence of differential 
cross sections,
the $W$ dependence of the total cross section,
and the ratio of cross sections for $\eta \pi^0$ to $\pi^0 \pi^0$
production.
Finally, Section~\ref{sec-6} summarizes the results and 
presents the conclusion of this paper.
 
\section{Experimental apparatus and event selection}
\label{sec-2}
Events consisting only of neutral final states are extracted from the data 
collected in the Belle experiment.
In this section, the Belle detector and event selection procedure are
described.

\subsection{Experimental apparatus}
A comprehensive description of the Belle detector is
given elsewhere~\cite{belle}.
We mention here only those
detector components that are essential for the present measurement.
Charged tracks are reconstructed from hit information in 
the silicon vertex detector and the central
drift chamber (CDC) located in a uniform 1.5~T solenoidal magnetic field.
The detector solenoid is oriented along the $z$ axis, which points
in the direction opposite to that of the positron beam. 
Photon detection and
energy measurements are performed with a CsI(Tl) electromagnetic
calorimeter (ECL).

For this all-neutral final state, we require that there be no
reconstructed tracks coming from the vicinity of
the nominal collision point. 
Therefore, the CDC is used for vetoing events with charged track(s). 
The photons from decays of the neutral pion and the $\eta$ meson
are detected and their momentum vectors are measured by the ECL. 
The ECL is also used to trigger signal events.
Two kinds of the ECL trigger
are used to select events of interest: the ECL total energy deposit
in the triggerable acceptance region (see the next subsection)
is greater than 1.15~GeV (the ``HiE'' trigger), or
the number of ECL clusters counted according
to the energy threshold at 110~MeV for segments
of the ECL is four or larger (the ``Clst4'' trigger).
The above energy thresholds are determined by studying the
correlations between the two triggers in the experimental data.
No software filtering is applied for triggering
events by either or both of the two ECL triggers.

\subsection{Experimental data and data filtering}
We use a 223~fb$^{-1}$ data sample from the Belle experiment
 at the KEKB asymmetric-energy $e^+e^-$ collider~\cite{kekb}.
The data were recorded at several $e^+e^-$ c.m. energies 
summarized in Table~\ref{tab:lum_data}.
The difference of the luminosity functions
(two-photon flux per $e^+e^-$-beam luminosity)
in the measured $W$ regions due to the difference of
the beam energies is small (maximum $\pm$ 4\%). 
We combine the results from the different beam energies. 
The effect on the cross section is less than 0.5\%.
\begin{center}
\begin{table}
\caption{Data sample: luminosities and energies}
\label{tab:lum_data}
\begin{tabular}{ccl} \hline \hline
~~$e^+e^-$ c.m. energy~~ & ~~~~Luminosity~~~~ & Runs \\
(GeV)& (fb$^{-1}$) & \\ \hline
10.58 & 179 & $\Upsilon(4S)$ \\ 
10.52 & 19 & continuum\\
10.36 & 2.9 & $\Upsilon(3S)$ \\
10.30 & 0.3 & continuum\\
10.86 & 21.7 & $\Upsilon(5S)$ \\ \hline
Total & 223 & \\
\hline\hline
\end{tabular}
\end{table}
\end{center}

The analysis is carried out in the ``zero-tag'' mode, where
neither the recoil electron nor positron are detected. 
We restrict the virtuality of the incident photons to be small
by imposing a strict requirement on the transverse-momentum balance with respect to 
the beam axis for the final-state hadronic system.

The filtered data sample (``Neutral Skim'') used for this analysis
is the same as the one used for
$\pi^0 \pi^0$ studies~\cite{pi0pi0,pi0pi02}. 
The important criteria in this filtering are:
no good tracks; two or more photons or one or more
neutral pions that satisfy a specified energy or 
transverse-momentum criterion. 
Performance of the ECL
triggers is studied in detail using the $\pi^0 \pi^0$ events
~\cite{pi0pi0}.

\subsection{Event selection}
From the Neutral Skim event sample,
we select $\gamma \gamma \to
\eta \pi^0$ with the following conditions: 
\newcounter{num}
\begin{list}%
{(\arabic{num})}{\usecounter{num}}
\setlength{\rightmargin}{\leftmargin}
\setlength{\topsep}{0mm}
\setlength{\parskip}{0mm}
\setlength{\parsep}{0mm}
\setlength{\itemsep}{0mm}
\item the total energy deposit in ECL
is smaller than 5.7~GeV;
\item  there are exactly four photons in the ECL each having energy
larger than 100~MeV;
\item  the ECL energy sum within the triggerable region
is larger than 1.25~GeV, or, all the four photons
are within the triggerable region, i.e.
in the polar-angle range, $-0.6255 < \cos \theta <+0.9563$,
in the laboratory frame;
\item  a combination of two photons is reconstructed as a neutral
pion that satisfies the following conditions on invariant mass,
$|M(\gamma \gamma) - 0.1350~\GeV/c^2|<0.0200~\GeV/c^2$, 
transverse momentum $p_t(\pi^0) > 0.15~\GeV/c$ and
goodness of the mass-constrained fit $\chi^2<9$;
\item  the combination of the remaining two photons has an invariant
mass consistent with $\eta \to \gamma \gamma$, 
$0.51~\GeV/c^2 < M(\gamma \gamma) < 0.57~\GeV/c^2$.
There are three combinations of photon pairs that can be constructed from
the four photons, and
all the combinations are tried and any of them satisfying the above criteria
are retained.
We scale the energy of the two photons in (5) with a factor
that is the ratio of the nominal $\eta$ mass to the 
reconstructed mass. 
This is equivalent to an approximate 1-C mass constraint 
fit where the relative energy resolution ($\Delta E/E$) is independent of $E$
and the resolution in the angle measurement is much better than
that of the energy. 
The present case is close to this.
After scaling the $\eta$'s four-momentum,
we calculate the invariant mass ($W$) and the transverse 
momentum ($|\Sigma {\vec p}_t^*|$)
in the $e^+e^-$ c.m. frame for the $\eta \pi^0$ system.
\item The transverse momentum
is  is required to be less than $0.05~\GeV/c$.
\end{list}

We define the c.m. scattering angle, $\theta^*$,
as the scattering angle of the $\pi^0$ (or equivalently,
of the $\eta$ )
in the $\gamma \gamma$ c.m. frame, for each event. 
We use an approximation that the $e^+e^-$ axis is the 
reference for this polar angle as
(since we do not know the exact $\gamma\gamma$ axis
in the zero-tag condition).

Signal and background events for $e^+e^- \to e^+e^- \eta\pi^0$ are
generated using the TREPS code~\cite{treps}.
All Monte Carlo (MC) events are put through
the trigger and detector simulators and the event selection program.
We find that up to 2\% of events in
the region below $W<1.05~\GeV$ have two entries per event 
because of the multiple combinations satisfying
criteria (4), (5) and (6). 
The two entries per
event have similar $W$ and $|\Sigma {\vec p}_t^*|$, but
different $|\cos \theta^*|$ values. 
The fraction with double entries is small and is in principle 
compensated by the 
normalization using the efficiency determined by the MC sample.
We find a similar fraction of 
multiple entries in the signal-MC data.

A total of $2.82 \times 10^5$ events are selected from
$3.53 \times 10^8$ events of the Neutral Skim sample.
The lego plots of two-dimensional distributions of the 
selected events (after requiring $|\Sigma {\vec p}_t^*|< 50$~MeV/$c$)
are shown in Fig.~\ref{fig01}. 
The projected $W$ distribution integrated over $|\cos \theta^*|<0.8$
is shown in Fig.~\ref{fig02}.
We find at least three resonant
structures: near 0.98~GeV ($a_0(980)$), 1.32~GeV ($a_2(1320)$) and
1.7~GeV (probably the $a_2(1700)$).

\begin{figure}
\centering
\includegraphics[width=13cm]{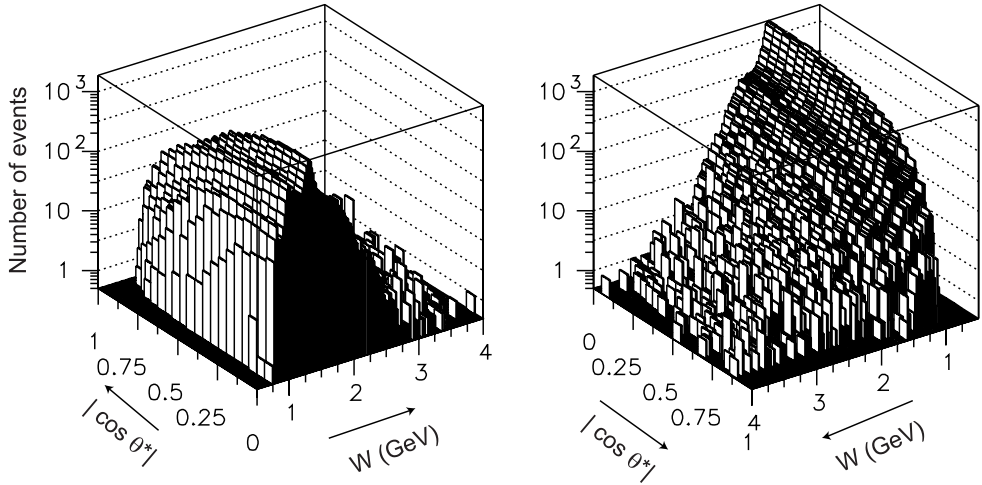}
\caption{Two-dimensional $W$ and $|\cos \theta^*|$ distribution 
of the $\eta \pi^0$ candidates in data.
The same distribution is viewed from two different directions.}
\label{fig01}
\end{figure}

\begin{figure}
\centering
\includegraphics[width=10cm]{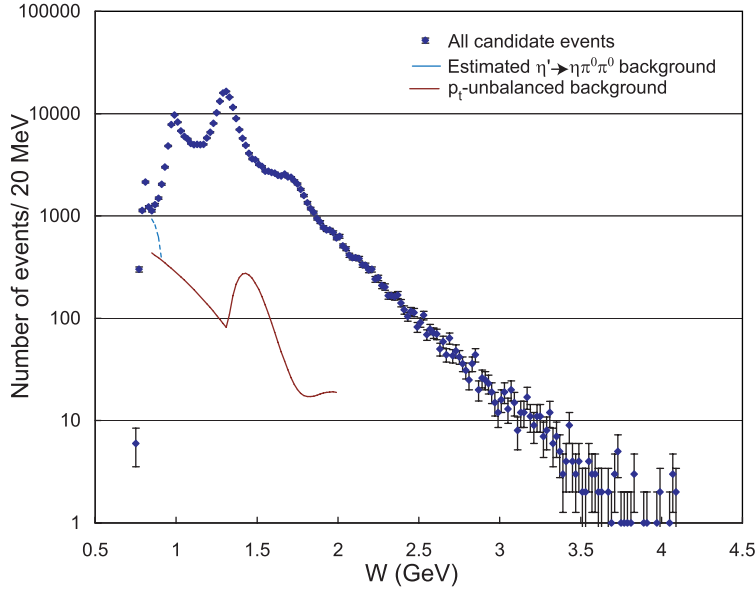}
\centering
\caption{$W$ distribution for $\eta \pi^0$ candidate events 
integrated over $|\cos \theta^*|<0.8$.
The solid curve is the $p_t$-unbalanced background that are 
experimentally determined. 
The dashed curve includes
an additional contribution from the $\eta' \to
\eta \pi^0 \pi^0$ background.}
\label{fig02}
\end{figure}

\section{Deriving differential cross sections}
\label{sec-3}
In this section, we present the procedure to derive differential cross 
sections.
First, the nature and origin of backgrounds and the method for
their subtraction are described.
Unfolding is then applied, efficiencies are determined,
differential cross sections are derived and their systematic errors are 
estimated.

\subsection{Background subtraction}
In the entire energy range the background is primarily from photons 
originated from spent electrons
as identified by unbalanced $p_t$. 
At low energy there is
additional background from the $\eta'$ decaying into $\eta\pi^0\pi^0$.

\subsubsection{Background from $\eta' \to \eta \pi^0 \pi^0$}
A scatter plot of $p_t$ balance 
vs. invariant mass for the candidate events, Fig.~\ref{fig03}(a), 
shows a concentration of events 
in the $p_t$-unbalanced region in the vicinity of 0.82~GeV. 
This energy is close to the mass
difference between the $\eta'$ and $\pi^0$, and this structure is 
due to the background from $\eta' \to \eta \pi^0 \pi^0$ where
two photons from one $\pi^0$ are undetected. 
Since the background
is larger than the signal in the $\eta'$ mass region, we cannot measure
the cross section below $W<0.84~\GeV$.
Kinematically, for $\eta'$ decays the invariant mass
of the detected $\eta \pi^0$ system cannot be
greater than  $0.823~\GeV/c^2$.
However, there is a rather long tail on the
higher side up to $W \sim 0.88~\GeV$. 
This background is due to
a fake pion reconstructed from a photon with another photon
from a $\pi^0$ or a noise cluster.
Figure \ref{fig03}(b) shows that $\eta$ mesons are relatively
cleanly reconstructed.
  
 We subtract the background from primary $\eta'$s in two-photon collisions  
using MC.
The normalization of the 
$\eta' \to  \eta \pi^0 \pi^0$ background is determined using the 
experimental data in the
control region, $0.80~\GeV <W< 0.84~\GeV$ integrated over all
angles and the transverse-momentum range, 
$|\Sigma {\vec p}_t^*|< 0.15~\GeV/c$ (Fig.~\ref{fig04}(a)). 
We assume that the signal yield is negligibly small in this $W$ range.
The $p_t$ distribution
is decomposed into the $\eta'$ background peaking near $0.05~\GeV/c$
and the other $p_t$-unbalanced component
by performing a fit, where we
use the signal and background functional shapes described in the next
subsection for the $\eta'$ and the latter background, respectively. 
The shape parameters of the $\eta'$ component are fixed to those obtained
from the fit to the corresponding MC data.
 
From the product of the two-photon decay width
and the branching fraction of the $\eta'$, 
$\Gamma_{\gamma\gamma}(\eta'){\cal B}(\eta' \to \eta \pi^0 \pi^0)$, 
we can estimate the absolute size of the background yield from this source. 
The ratio of the observed background to the MC expectation
is $0.91 \pm 0.02(stat.) \pm 0.09(syst.) \pm
0.06 (\Gamma_{\gamma\gamma}{\cal B})$, where the 
first two errors are 
experimental, and the last error is from the uncertainty of the
known $\eta'$ properties. 
This factor is consistent with unity.

Using the normalization thus determined and the MC events, the 
background yields from this source in each angular bin in the
range $0.84~\GeV<W<0.90~\GeV$ are determined. 
This background component is incorporated in the fit described in the 
next subsection with the yield and shape fixed. 
We neglect the $\eta'$ background in the $W$ region above 0.90~GeV.

\begin{figure}
\centering
\includegraphics[width=13cm]{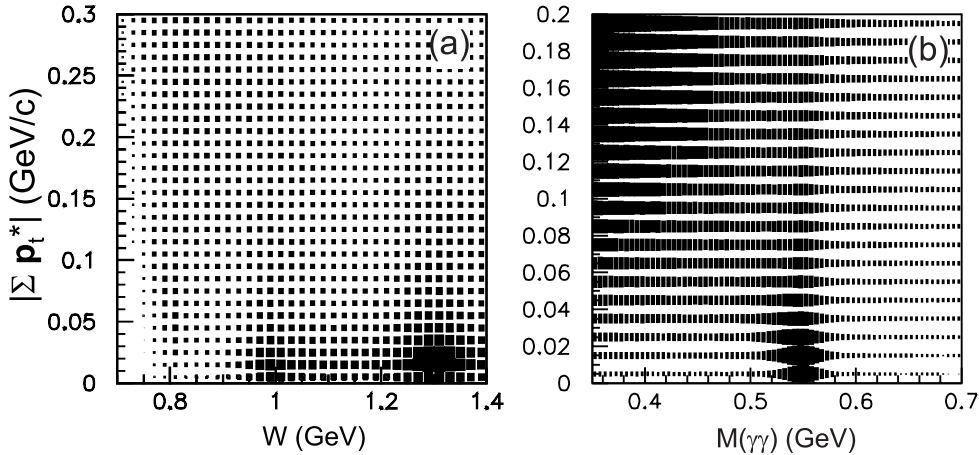}
\caption{(a)Two-dimensional ($W$, $|\sum {\vec p}_t^*|$)
distribution of the $\eta \pi^0$ candidate events in the relatively
low-$W$ region. 
(b)Two-dimensional distribution of the invariant
mass of the two detected photons near the $\eta$ mass and
the $p_t$-balance of events calculated from the mass of
four photons (before applying the mass constraints) for events with
$M(\eta \pi^0) < 1.0~\GeV/c^2$.}
\label{fig03}
\end{figure}

\begin{figure}
\centering
\includegraphics[width=10cm]{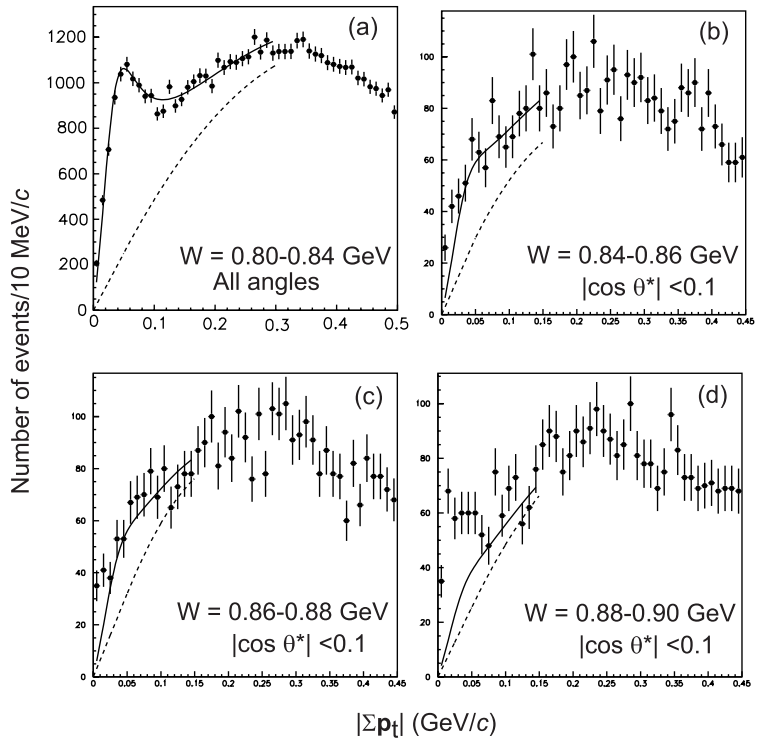}
\centering
\caption{(a): The $p_t$ distribution of the $\eta \pi^0$ candidates
at $W=0.80$-0.84~GeV integrated over $|\cos \theta^*|<0.8$.
The peak near $0.05~\GeV/c$ is attributed
to the background from $\eta' \to \eta \pi^0 \pi^0$ and
the normalization is determined by a fit to MC events of this process. 
The dashed curve shows
the experimentally determined background from the $p_t$-unbalanced
background component and the solid curve is the sum of
the two background components.
(b), (c) and (d): The $p_t$ distribution 
of the $\eta \pi^0$ candidate events in the shown kinematical regions.
Estimated background yields of the two components are shown by the
curves (explained in the caption for (a)).  
The excess over the solid 
curve near $|\sum {\vec p}_t^*|=0$ corresponds to the signal process.}
\label{fig04}
\end{figure}

\subsubsection{Subtraction of backgrounds}
In the $W$ region below 2.0~GeV, the $p_t$-unbalanced component
is non-negligible and is subtracted by fitting the $p_t$
distributions. 
The fitting function is a sum of the signal and background components.
The $\eta'$ background estimated in the previous subsection
is incorporated in the fit with a fixed shape and size
for $W<0.90$~GeV (Figs.~\ref{fig04}(b-d)).

The signal component follows an empirical parameterization
from signal MC:
\begin{equation}
y=\frac{Ax}{x^{2.1}+B+Cx} \; ,
\label{eqn:backg}
\end{equation}
($x$ is  $|\sum {\vec p}_t^*|$, $A$, $B$ and $C$ are
fitting parameters), 
where the distribution has a linear shape near
$x=0$ and decreases as $\sim x^{-1.1}$ at large $x$. 
Here $B$ is an important parameter that determines the peak position;
the peak is at $x=B^{\frac{1}{2.1}}/1.1$. 
The background is parameterized by
a linear function vanishing at $x=0$ in $x<50$~MeV/$c$
and a second-order polynomial for  $x>50$~MeV/$c$ connected
smoothly (up to the first derivative) at $x=50$~MeV/$c$.
Fits are applied for $|\Sigma {\vec p}_t^*|<200$~MeV/$c$ in each
bin of $W < 2.0$~GeV and $|\cos \theta^*| <0.8$ with the 
bin widths of 0.04~GeV and 0.1 for the two directions, respectively.

The background yields found from the fits 
are fitted to a smooth two-dimensional
function of ($W$, $|\cos \theta^*|$), in order to minimize
the statistical fluctuations from the MC simulation.
In Fig.~\ref{fig02}, the curves corresponding to the background
thus determined are shown, as well as the fixed background
from $\eta'$ decays.

 The backgrounds are subtracted from the experimental yield distribution.
The $\eta'$ background estimated in the previous subsection
is also subtracted at $W<0.90~\GeV$.  
The error arising from
this background subtraction is taken into account as a systematic error
(see the subsection E).

To confirm the validity of the background subtraction method,
we examine the $W$ dependence of the yield ratio $R$ 
defined as:
\begin{equation}
R= \frac{Y(0.15~\GeV/c < |\sum \vec{p}_t^*| < 0.20~\GeV/c)}
  {Y (|\sum \vec{p}_t^*|<0.05~ \GeV/c)} \; ,
\label{eqn:r_pt}
\end{equation}
where $Y$ is the yield in the specified region.
We observe that structures have similar features as those seen in the
estimated $p_t$-unbalanced background (Fig.~\ref{fig05}).
There is a peaking structure in 
the background around $W=1.5$~GeV, and a small enhancement
just above 2.0~GeV. 
No significant structures are seen from 2.5~GeV up to $\sim 3.3$~GeV. 

We observe a significant increase of $R$ above 3.3~GeV. 
Such an increase
is not reproduced by the MC simulation, where only a slow increase of $R$
($R=0.15$ at 2.0~GeV and $R=0.25$ at 4.0~GeV) is expected. 
The excess of $R$, $\Delta R$, is the contribution from 
multibody background processes.
We apply a background 
subtraction with the background contamination level estimated to be
$\Delta R/6$.
This factor is obtained by assuming a quasi-linear $|\sum \vec{p}_t^*|$
dependence of the background and extracting its leakage into the
signal region ($|\sum \vec{p}_t^*|<0.05~\GeV/c$), which is approximately
1/6 of the
yield in the $0.15~\GeV/c < |\sum \vec{p}_t^*|<0.20~\GeV/c$ region.
We take a half of the correction (i.e. $\Delta R/12$)
as the systematic error from this source if this systematic is larger than
the 3\% uncertainty nominally applied for the whole region above 2.0~GeV 
(see the section on systematic errors). 
The actual sizes of the correction used (i.e. $\Delta R/6$) in $W$ bins are
3.8\% (3.35~GeV), 
8.5\% (3.65~GeV),
and 13.0\% (3.95~GeV), respectively. 
These corrections are much smaller than the statistical errors in
these $W$ bins.

We find that the background in the $\eta$ mass sideband is negligibly small
after the subtraction of these backgrounds; there are no $p_t$-balanced 
backgrounds in the $\eta$ mass sideband of the $p_t$-balance distribution
shown in Fig.~\ref{fig03}(b). 
Therefore, we do not perform background subtraction for the non-$\eta$ 
component.

\begin{figure}
\centering
\includegraphics[width=9cm]{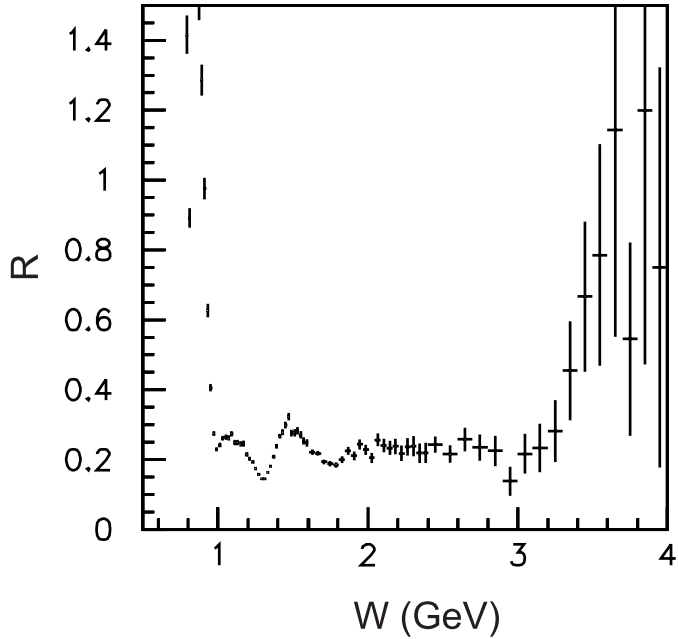}
\centering
\caption{Energy dependence of the ratio of the yield in 
the $p_t$-unbalanced region to that in the balanced region 
(see text for the exact definition), which indicates the
level of background contamination.}
\label{fig05}
\end{figure}

\subsection{Unfolding the $W$ distributions}
We unfold the experimental yields
in the $W$ distributions to correct for the finite invariant-mass
resolution in the measurement. 
The smearing function
is an asymmetric Gaussian function, which is determined 
by the signal MC with a further empirical correction.
The standard deviations of the function are assumed to follow 
$1.3 \times (1.4-0.3/W^2)$\% ($W$ is in GeV, and the resolution varies 
by 1.3\%-1.8\% depending on $W$)
on the lower side of the peak and 
$0.77 \times (1.4-0.3/W^2)$\% (varies by 0.8\%-1.1\%) 
on the higher side. 
The resolution is slightly better than that in the $\pi^0\pi^0$ case
in the low $W$ region, because of the large opening angle of the two photons
from $\eta$ decay.

The unfolding procedure is applied for 0.9~GeV $< W <$ 1.6~GeV with
a bin width of 0.02~GeV, and  for 1.6~GeV $ < W < $ 2.4~GeV with
a bin width of 0.04~GeV. 
The unfolding is done independently
in each angular bin, whose width is $\Delta |\cos \theta^*| = 0.05$
for $W<1.6$~GeV and  $\Delta |\cos \theta^*| = 0.1$
for $W>1.6$~GeV. 
Figure~\ref{fig06} shows the yield distributions
before and after the unfolding in the smallest $|\cos \theta^*|$ bins. 
At higher energies, no unfolding is applied since the experimental
yield is still insufficient.

\begin{figure}
\centering
\includegraphics[width=12cm]{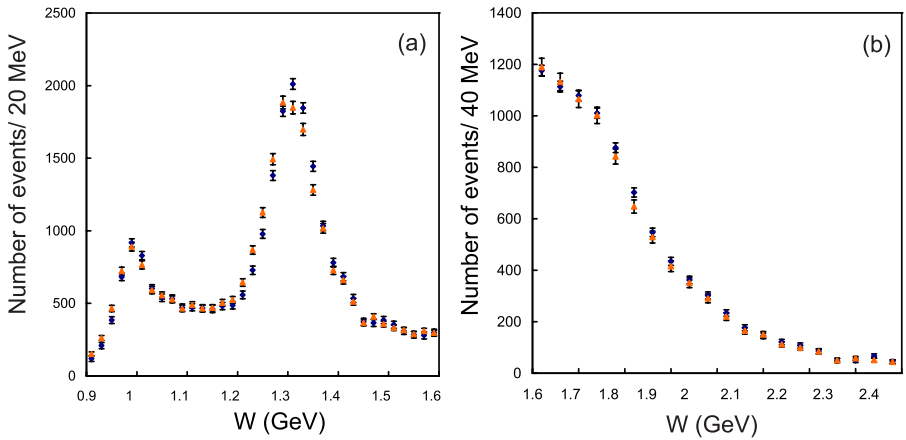}
\centering
\caption{ The yield distributions
before (triangles) and after (diamonds) the 
unfolding in the smallest 
$|\cos \theta^*|$ bins, (a) $|\cos \theta^*|<0.05$ and  
 (b) $|\cos \theta^*|<0.1$.} 
\label{fig06}
\end{figure}

\subsection{Determination of the efficiency}
 The signal MC events for $e^+e^- \to e^+e^- \eta\pi^0$ are
generated using the TREPS code~\cite{treps} for
the efficiency calculation at 36 fixed $W$ points between 0.75 and 4.2~GeV
and isotropically in $|\cos \theta^*|$. 
The angular distribution at 
the generator level does not play a role in the efficiency
determination, because we calculate the efficiencies separately 
in each $|\cos \theta^*|$ bin with a width of 0.05.
The $Q^2_{\rm max}$ parameter
that gives a maximum virtuality of the incident
photons is set to 1.0~GeV$^2$, and the form factor 
for the cross sections for the virtual photon collisions,
$\sigma_{\gamma\gamma}(0,Q^2)=\sigma_{\gamma\gamma}(0,0)/
(1+Q^2/W^2)^2$ is used. 
This form factor does not
play any essential role in the present analysis, since
our stringent $p_t$-balance cut
($|\sum \vec{p}_t^*| < 0.05~\GeV/c$) requires $Q^2/W^2$ for the 
selected events to be much smaller than 1.
A sample of 400,000 events is 
generated at each $W$ point and is subjected to
the detector and trigger simulations. 
The obtained efficiencies are fitted to a two-dimensional function
of ($W$, $|\cos \theta^*|$) with an empirical functional form.

We embed background hit patterns from random trigger data
into MC events. 
We find that different samples of background
hits give small variations in the selection efficiency determination.
A $W$-dependent error in the efficiency, 2-4\%, arises from 
the uncertainty in this effect.
Figure~\ref{fig07} shows the two-dimensional dependence 
of the efficiency on ( $W$, $|\cos \theta^*|$) 
after the fit for smoothing. 

\begin{figure}
\centering
\includegraphics[width=9cm]{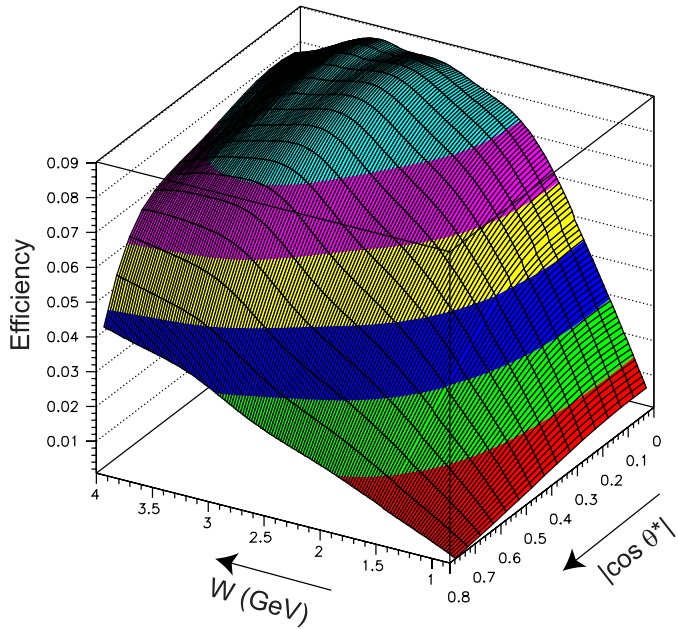}
\centering
\caption{ Two-dimensional dependence of the efficiency
on ($W$, $|\cos \theta^*|$).}
\label{fig07}
\end{figure}

\subsection{Derivation of differential cross sections}
The differential cross section for each
($W$, $|\cos \theta^*|$) point is given by:
\begin{equation}
\frac{d\sigma}{d|\cos \theta^*|} =
\frac{\Delta Y - \Delta B}{\Delta W \Delta |\cos \theta^*| 
\int{\cal L}dt L_{\gamma\gamma}(W)  \eta } \; ,
\label{eqn:dcs}
\end{equation}
where $\Delta Y$ and $\Delta B$ are the signal yield and
the estimated $p_t$-unbalanced background in the bin, 
$\Delta W$ and $\Delta |\cos \theta^*|$ are the bin widths, 
$\int{\cal L}dt$ and  $L_{\gamma\gamma}(W)$ are
the integrated luminosity and two-photon luminosity function
calculated by TREPS~\cite{treps}, respectively, and  $\eta$ is the  
efficiency including the correction described in the previous section.
The bin sizes for $W$ and $\Delta |\cos \theta^*|$ are summarized in
Table~\ref{tab:binsize}.
\begin{center}
\begin{table}
\caption{Bin sizes}
\label{tab:binsize}
\begin{tabular}{lcc} \hline \hline
$W$ range & $\Delta W$ & $\Delta |\cos \theta^*|$ \\ 
(GeV) & (GeV) & \\\hline
0.84 -- 1.6 & 0.02 & 0.05 \\
1.6 -- 2.4 & 0.04 & 0.10 \\
2.4 -- 4.0 & 0.10 & 0.10 \\
\hline\hline
\end{tabular}
\end{table}
\end{center}

Figure~\ref{fig08} shows the angular dependence of the 
differential cross sections for some selected $W$ regions. 
Figures~\ref{fig09} show the cross section integrated 
over $|\cos \theta^*|<0.8$ on logarithmic and linear
scales for partial $W$ regions.
The data points are in good agreement with those of Crystal Ball~\cite{cbep}.

\begin{figure}
\centering
\includegraphics[width=14cm]{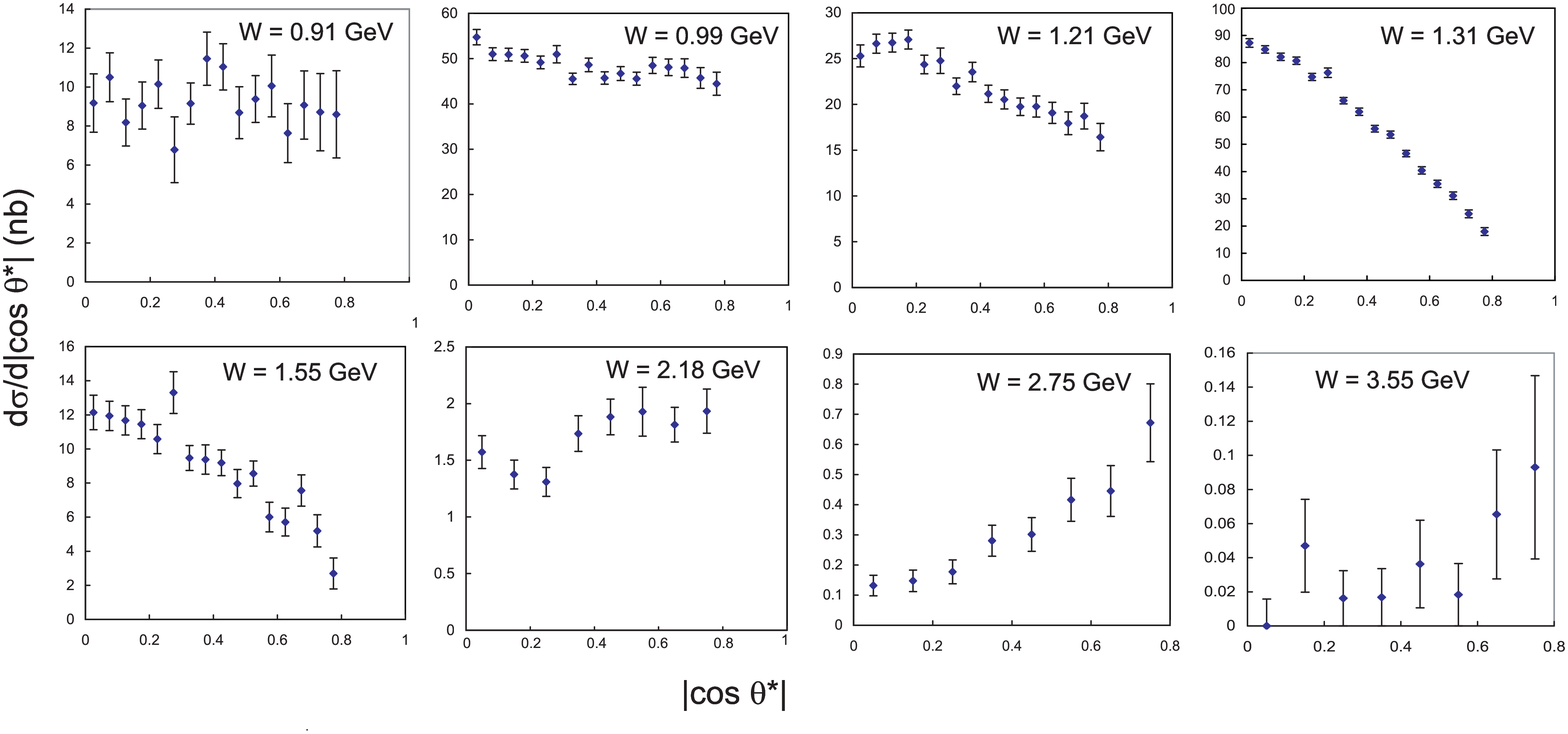}
\centering
\caption{Angular dependence of the 
differential cross sections for eight selected $W$ bins indicated.
The bin sizes are summarized in Table~\ref{tab:binsize}.} 
\label{fig08}
\end{figure}

\begin{figure}
\centering
\includegraphics[width=10cm]{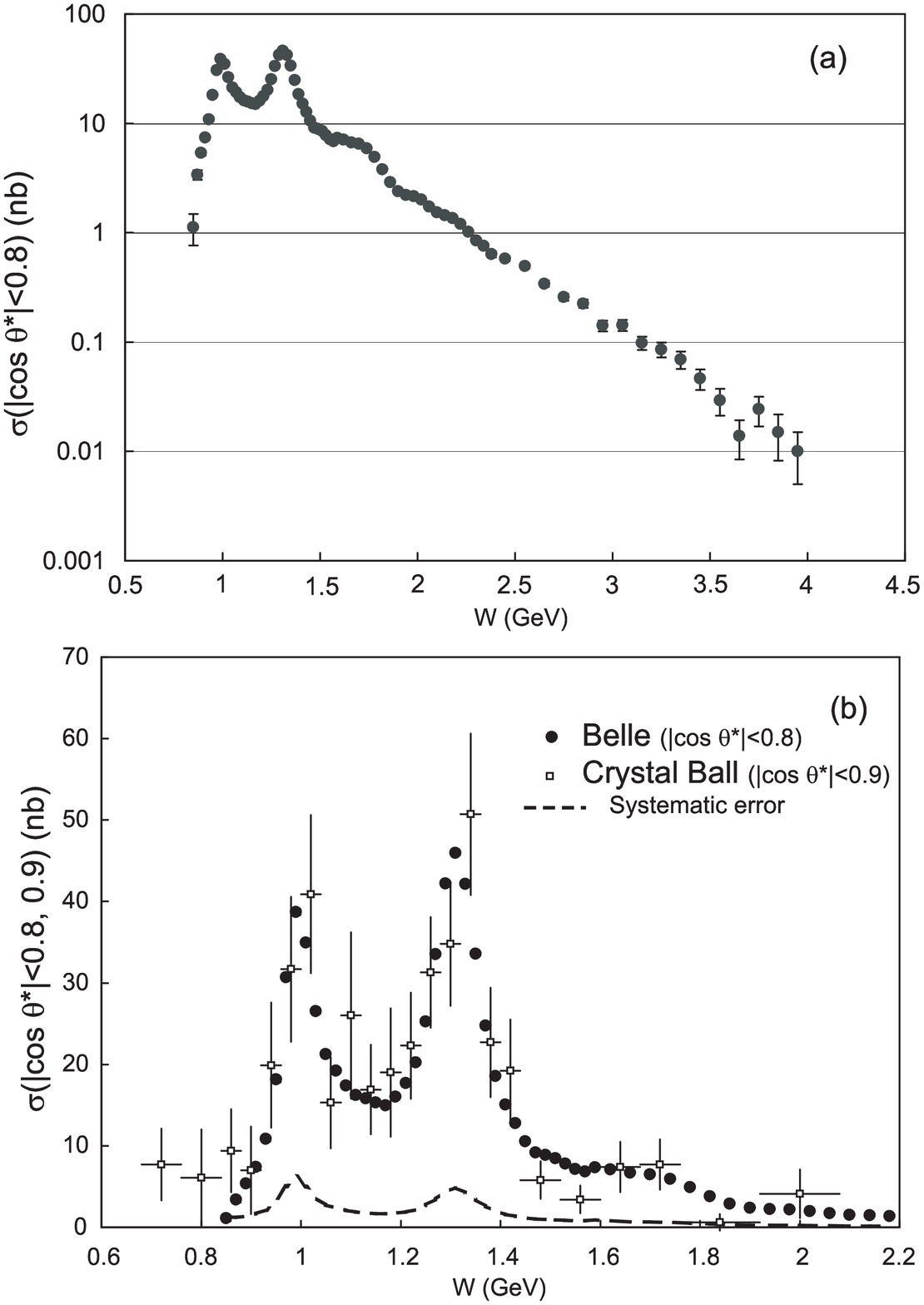}
\centering
\caption{The cross section integrated 
over $|\cos \theta^*|<0.8$ on a logarithmic (a) and linear
(b) scale compared with the Crystal Ball measurement
($|\cos \theta^*|<0.9$)~\cite{cbep}.
The corrections for different $|\cos \theta^*|$ coverage are not made.
The dashed curve shows the size of the systematic error.} 
\label{fig09}
\end{figure}

\subsection{Systematic errors}
Various sources of systematic uncertainties assigned
for the $\eta \pi^0$ signal yield, efficiency and the cross section
evaluation are described in detail below
and summarized in Table~\ref{tab:sysdcs}.
 
{\it Trigger efficiency}: The systematic error due to  
the Clst4 trigger (the mnemonics of the ECL triggers were described in
Sec.~II.A) is taken to be $2/3$ of the difference
in the efficiencies
when the thresholds for the Clst4 trigger are varied
from 110~MeV to 100~MeV in the trigger simulator, and 
no error is assigned for $W>2.5$~GeV where the HiE trigger
plays a dominant role.
In addition, we take the uncertainty in the efficiency of the 
HiE trigger to be 4\% in the whole $W$ region. 
The systematic
errors from the two triggers are combined in quadrature.
The former component is approximated by an angular-independent
function of the c.m. energy, $27\% \times 0.5^{10(W-0.85~[{\rm GeV}])}$.
This exceeds 5\% for $W<1.2$~GeV.

{\it The reconstruction efficiency}: we assign 6\% for the reconstruction of
a $\pi^0$ and an $\eta$.

{\it $p_t$-balance cut}: 3\% is assigned. 
The $p_t$-balance distribution for the signal is well reproduced 
by MC so that the efficiency is correct to within this error.

{\it Background subtraction}: 20\% of the size of the 
subtracted component is assigned
to this source for the range $0.9~\GeV < W < 2.0~\GeV$.
In the $W$ region where the background subtraction 
is not applied ($W>2.0$~GeV), we assign a systematic error 
of 3\%, which is a conservative upper limit on the background
contamination from an investigation of the experimental distributions. 
Above 3.5~GeV, the error originating from the background subtraction 
($\Delta R/12$) is larger than 3\%, and we replace the error by the latter 
value.
In the region $0.84~\GeV < W < 0.90~\GeV$ where the
$\eta'$ background is subtracted, the assigned error is 
a quadratic sum of 
7\% of the subtracted $\eta'$ background and  20\% of 
the subtracted $p_t$-unbalanced background.

{\it Luminosity function}: We assign 4\% (5\%) for $W< \; (>) \; 3.0$~GeV.

{\it Beam background effect for event selection}: We assign a 2\% - 6\% error
depending on $W$ for uncertainties of the inefficiency in event selection 
due to beam-background photons. 
The uncertainty is estimated from
the variation of efficiencies among different experimental periods or
background conditions.
We adopt the averaged efficiency from the
different background files, and the uncertainty in the average
is assigned as the error.

{\it Unfolding}: 
Uncertainties from the unfolding procedure, using the single value
decomposition approach in Ref.~\cite{svdunf},
are estimated by varying the effective-rank parameter
of the decomposition within reasonable bounds.

{\it Other efficiency errors}: An error of 4\% is assigned for
uncertainties in the efficiency determination based on MC including
the smoothing procedure.

 The total systematic error is obtained by adding all the sources in quadrature
and is 10-12\% for the intermediate and high $W$ regions. 
It becomes much larger for $W < 1.06~\GeV$.

\begin{center}
\begin{table}
\caption{Systematic errors for the differential cross sections.
Ranges of errors are shown when they depend on $W$.}
\label{tab:sysdcs}
\begin{tabular}{lc} \hline \hline
Source & Error (\%) \\ \hline
Trigger efficiency & 4 -- 30\\
$\eta$ and $\pi^0$ reconstruction efficiency & 6 \\
$p_t$-balance cut  & 3 \\
Background subtraction & 3 -- 35 \\
Luminosity function & 4 -- 5 \\
Overlapping hits from beam background & 2 -- 6 \\ 
Unfolding procedure & 0 -- 4 \\
Other efficiency errors & 4 \\ \hline
Overall  & 10 -- 12 (for $W > 1.06~\GeV$)\\
\hline\hline
\end{tabular}
\end{table}
\end{center}

\section{Study of resonances}
\label{sec-4}
In this section, we extract the resonance parameters of
the $a_0(980)$ and a possible resonance $a_0(1450)$,
as well as check the consistency of the $a_2(1320)$ parameters.
We also study whether or not the $a_2(1700)$ is produced in this
reaction.

\subsection{Formalism}
The formalism is exactly the same as that for $\pi \pi$~\cite{mori1,
mori2, pi0pi0, pi0pi02} and
the analysis is quite similar.
In the energy region $W \leq 2.0~\GeV$, $J > 2$ partial waves (the next is 
$J=4$) may be neglected so that only S and D waves are to be considered.
The differential cross section can be expressed as:
\begin{equation}
\frac{d \sigma}{d \Omega} (\gamma \gamma \to \eta \pi^0)
 = \left| S \: Y^0_0 + D_0 \: Y^0_2  \right|^2 
+ \left| D_2 \: Y^2_2  \right|^2 \; ,
\label{eqn:diff}
\end{equation}
where $S$ represents the S wave, $D_0$ ($D_2$) denotes the helicity 0 (2) 
components of the D wave, respectively, and $Y^m_J$ are the spherical 
harmonics.
Since the $|Y^m_J|$'s are not independent of each other,
partial waves cannot be separated using measurements of
differential cross sections alone.
To overcome this problem, we write Eq.~(\ref{eqn:diff}) as:
\begin{equation}
\frac{d \sigma}{4 \pi d |\cos \theta^*|} (\gamma \gamma \to \eta \pi^0)
 = \hat{S}^2 \: |Y^0_0|^2  + \hat{D}_0^2 \: |Y^0_2|^2
+ \hat{D}_2^2  \: |Y^2_2|^2 \, .
\label{eqn:diff2}
\end{equation}
The amplitudes $\hat{S}^2$, $\hat{D}_0^2$ and $\hat{D}_2^2$
correspond to the cases where interference terms are neglected; 
they can be expressed in terms of 
$S,~D_0$ and $D_2$ as follows~\cite{pi0pi0}:
\begin{eqnarray}
\hat{S}^2 &=& |S|^2 + \sqrt{5} {\rm Re}{(S^* D_0)} 
\; , \nonumber \\
\hat{D}_0^2 &=& |D_0|^2 + \frac{1}{\sqrt{5}} {\rm Re}{(S^* D_0)} 
\; , \nonumber \\
\hat{D}_2^2 &=& |D_2|^2 - \frac{6}{\sqrt{5}} {\rm Re}{(S^* D_0)} \; .
\label{eqn:def1}
\end{eqnarray}
Since squares of spherical harmonics are independent of each other,
we can fit the differential cross section at each $W$ to obtain 
$\hat{S}^2$, $\hat{D}_0^2$, and $\hat{D}_2^2$.
The unfolded differential cross sections are fitted
by taking into account statistical errors only,
which will not be independent at each $W$ because of the unfolding procedure.
However, we treat them as independent in the fit.
The resulting $\hat{S}^2$, $\hat{D}_0^2$ and $\hat{D}_2^2$ spectra
for $W < 2~\GeV$ are shown in Figs.~\ref{fig10}.
\begin{figure}
 \centering
   {\epsfig{file=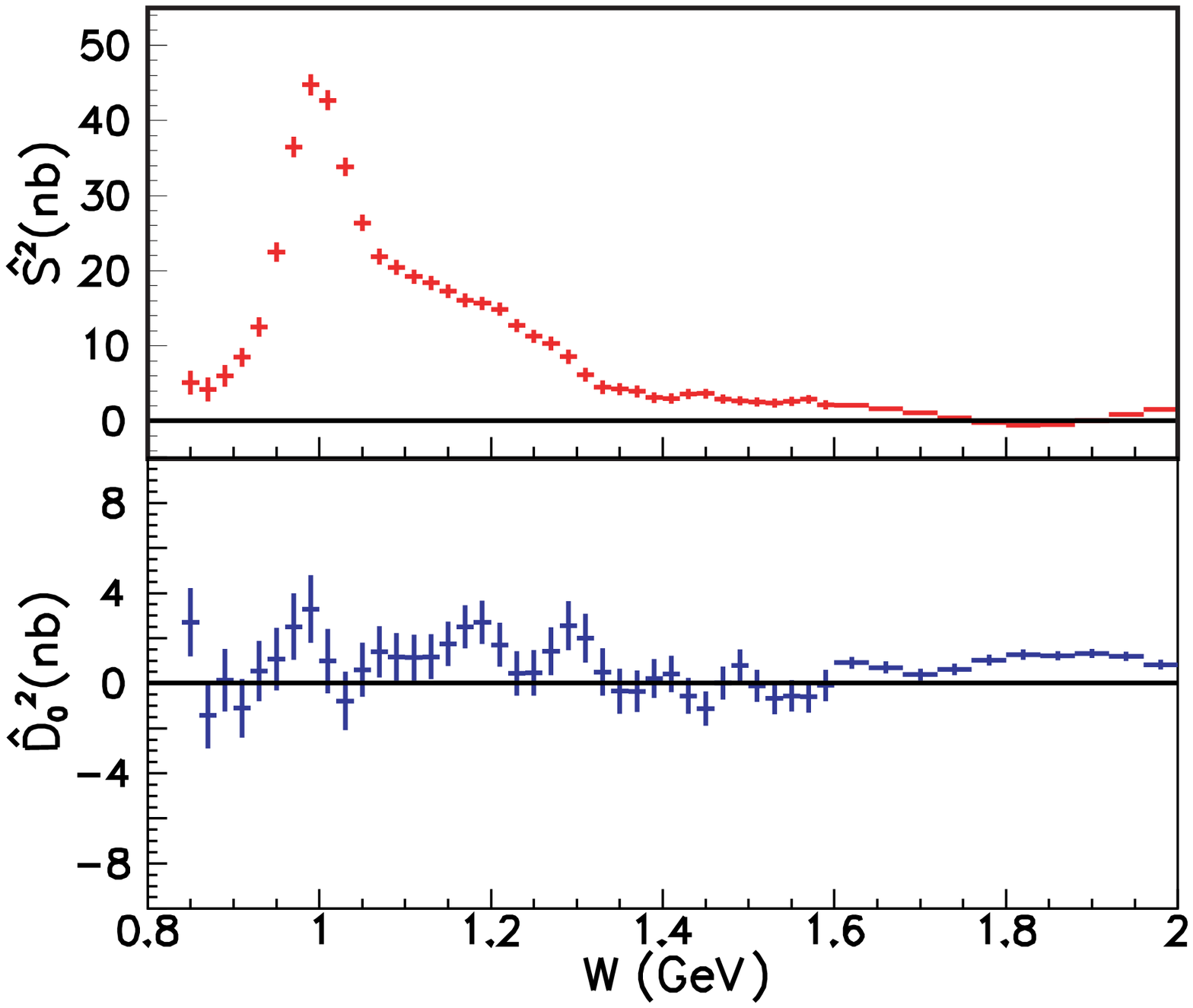,width=60mm}}
   {\epsfig{file=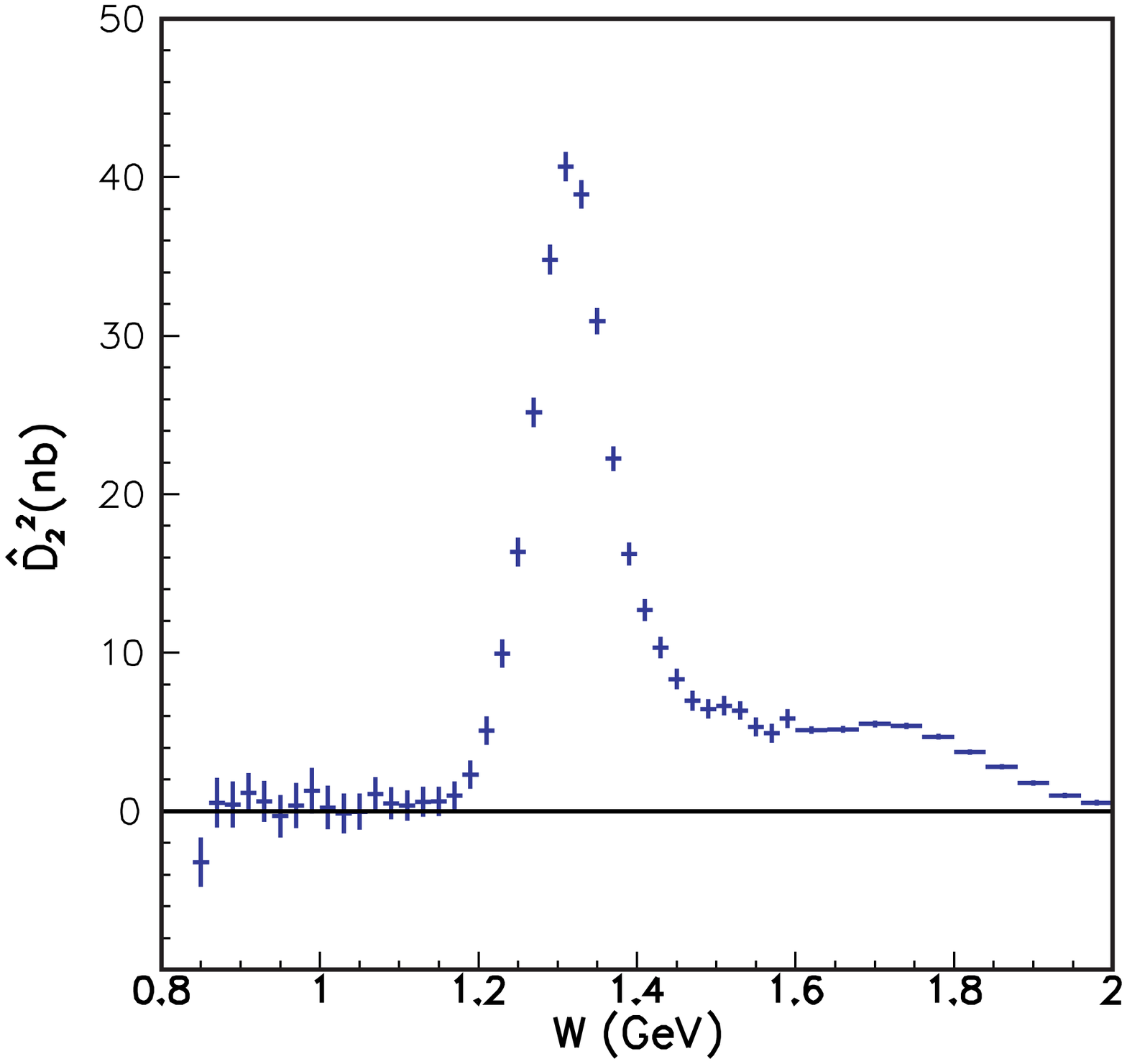,width=60mm}}
 \caption{Resulting spectra of $\hat{S}^2$, $\hat{D}_0^2$ and $\hat{D}_2^2$.
The error bars shown are diagonal statistical errors.}
\label{fig10}
\end{figure}

\subsection{Fitting partial wave amplitudes}
Although the derived amplitudes $\hat{S}^2$, $\hat{D}_0^2$ and $\hat{D}_2^2$
are functions of partial waves (Eq.~(\ref{eqn:def1})), they do give some 
indication of their behavior.
Notably, the D$_0$ wave appears to be small
and the D$_2$ wave is dominated by the $a_2(1320)$ with a hint of
the $a_2(1700)$.
The peak in $\hat{S}^2$ around $W=0.98~\GeV$ 
(Fig.~\ref{fig10}) is clearly due to the $a_0(980)$ resonance and
a shoulder above the $a_0(980)$ peak may be due to the $a_0(1450)$.

In this section, we derive information about resonances 
by parameterizing partial wave amplitudes and fitting differential 
cross sections.
Note that we do not fit the obtained $\hat{S}^2$, $\hat{D}_0^2$ and 
$\hat{D}_2^2$ spectra but fit the differential cross sections directly.
Once the functional forms of amplitudes are assumed, we can use 
Eq.~(\ref{eqn:diff}) to fit differential cross sections. 
We can then neglect the correlations between
$\hat{S}^2$, $\hat{D}_0^2$ and $\hat{D}_2^2$.
The $\hat{S}^2$, $\hat{D}_0^2$ and $\hat{D}_2^2$ spectra
are used to make an initial determination of which resonances are
important and to check fit quality.

The energy range of this measurement can be naturally divided 
into a low energy region
($W \leq 1.5~\GeV$), where the $a_0(980)$ and $a_2(1320)$ are important,
and a higher energy region ($W > 1.5~\GeV $), where the parameters of the
$a_2(1700)$ will be of interest.
In both regions, we neglect $J>2$ waves.

First we extract parameters of the $a_2(1700)$ in the high energy range.
A hint of the $a_2(1700)$ is visible in Fig.~\ref{fig09}.
Various fits have been performed: a fit in the energy region 
$1.5~\GeV < W < 2.0~\GeV$
with the parameters of the $a_2(1700)$ all floated or the mass and width fixed
to the PDG values; a fit in the energy region $0.9~\GeV < W < 2.0~\GeV$
by fixing the lower energy parameters to the ones determined below and
either floating or partially fixing the parameters of the $a_2(1700)$.
The resulting parameters of the $a_2(1700)$ vary much.
In addition, the fit quality is poor.
Thus, we cannot report a definite conclusion on the $a_2(1700)$.

In the low energy region, $W < 1.5~\GeV$, we can safely neglect $J>2$ waves.
The $\hat{D}_0^2$ contribution is small while $\hat{D}_2^2$ is 
seen to be dominated by the $a_2(1320)$ resonance.
We assume that the $a_2(1320)$ contributes to the $D_2$ wave only, 
since $\hat{D}_0^2$ is small.
The shoulder in the $\hat{S}^2$ spectrum
above the $a_0(980)$ peak may be due to the $a_0(1450)$.
However, in this fit we introduce a new resonance $a_0(Y)$ instead,
since its parameters 
are found to be quite different from those of the  $a_0(1450)$.
The goal of analysis is to obtain parameters of the $a_0(980)$,
$a_0(Y)$ and to check the consistency of the $a_2(1320)$ parameters that
have been measured well in the past.

\subsubsection{Parameterization of amplitudes}
We parameterize $S$, $D_0$ and $D_2$ waves as follows.
\begin{eqnarray}
S &=& A_{a_0(980)} e^{i \phi_{s0}} +  A_{a_0(Y)} e^{i \phi_{s1}} 
 + B_S \; , \nonumber \\
D_0 &=& B_{D0} \; , \nonumber \\
D_2 &=& A_{a_2(1320)} e^{i \phi_{d2}} + B_{D2} \; ,
\label{eqn:param}
\end{eqnarray}
where $A_{a_0(980)}$, $A_{a_0(Y)}$ and $A_{a_2(1320)}$ are the 
amplitudes of the $a_0(980)$, $a_0(Y)$ and $a_2(1320)$, respectively;
$B_S$, $B_{D0}$ and $B_{D2}$ are non-resonant 
(called hereafter ``background'')
amplitudes for S, D$_0$ and D$_2$ waves; and
$\phi_{s0}$, $\phi_{s1}$, and $\phi_{d2}$
are the phases of resonances relative to background amplitudes.
We also study the case with no $a_0(Y)$ and the case with the mass
of the $a_0(Y)$ fixed to that of the $a_0(1450)$.

The background amplitudes are parameterized as follows.
\begin{eqnarray} 
B_S &=&  a_s W'^2  + b_s W' + c_s
 + i (a'_s W'^2  + b'_s W'  + c'_s) , 
\nonumber \\
B_{D0} &=& a_0 W'^2  + b_0 W' + c_0
 + i (a'_0 W'^2  + b'_0 W'  + c'_0) , 
\nonumber \\
B_{D2} &=& a_2 W'^2  + b_2 W' + c_2
 + i (a'_2 W'^2  + b'_2 W'  + c'_2) .
\label{eqn:para2}
\end{eqnarray}
Here $W' = W-W_{\rm th}$ where $W_{\rm th}$ is the threshold energy.
We assume background amplitudes to be quadratic
in $W$ for the both real and imaginary parts of all waves.
In this way, symmetries among amplitudes are kept.
The arbitrary phases are fixed by choosing $\phi_{s0} = \phi_{d2} =0$.
We constrain all the background amplitudes to be zero at the threshold
by setting the $c$ and $c'$ parameters to zero 
in accordance with the
expectation that the cross section vanishes at the Thomson limit.

The relativistic Breit-Wigner resonance amplitude
$A_R(W)$ for a spin-$J$ resonance $R$ of mass $m_R$ is given by
\begin{eqnarray}
A_R^J(W) &=& \sqrt{\frac{8 \pi (2J+1) m_R}{W}} 
\frac{\sqrt{\Gamma_{\rm tot}(W) \Gamma_{\gamma \gamma}(W) {\B}(\eta \pi^0)}}
{m_R^2 - W^2 - i m_R \Gamma_{\rm tot}(W)} \; ,
\label{eqn:arj}
\end{eqnarray}
The energy-dependent total width $\Gamma_{\rm tot}(W)$ is given by
\begin{equation}
\Gamma_{\rm tot}(W) = \sum_X \Gamma_{X_1 X_2} (W) \; ,
\label{eqn:gamma}
\end{equation}
where $X_i$ is $\pi$, $K$, $\eta$, $\gamma$, etc.
For $J=2$ (the $a_2(1320)$ meson),
the partial width $\Gamma_{X_1 X_2}(W)$ is parameterized as~\cite{blat}:
\begin{equation}
\Gamma_{X_1 X_2} (W) = \Gamma_R {\cal B}(R \rightarrow X_1 X_2) 
\left( \frac{q_X(W^2)}{q_X(m_R^2)} \right)^5
\frac{D_2\left( q_X(W^2) r_R \right)}{D_2 \left( q_X(m_R^2) r_R \right)} \;,
\label{eqn:gamx}
\end{equation}
where $\Gamma_R$ is the total width at the resonance mass,
$q_X(W^2) = \sqrt{(W^2 - (m_{X_1} + m_{X_2})^2)
(W^2 - (m_{X_1} - m_{X_2})^2)}/(2W)$, 
$D_2(x) = 1/(9 + 3 x^2 +x^4)$,
and $r_R$ is an effective interaction radius that varies from 1~$\GeV^{-1}$ 
to 7~$\GeV^{-1}$ in different hadronic reactions~\cite{grayer}.
For the three-body and the other decay modes,
$\Gamma_{3-{\rm body}} (W) = \Gamma_R {\cal B}(R \rightarrow 3-{\rm body})
\frac{W^2}{m_R^2}$ is used instead of Eq.~(\ref{eqn:gamx}).
All the parameters of the $a_2(1320)$ are fixed to the PDG values
 as listed in Table~\ref{tab:a2para}~\cite{pdg},
except for $r_R$ which is fitted to be $3.09^{+0.53}_{-0.55}~(\GeV/c)^{-1}$,
consistent with $3.62 \pm 0.03~(\GeV/c)^{-1}$ determined 
for the $f_2(1270)$~\cite{mori2}.
For the $a_0(980)$ and $a_0(Y)$, the widths are taken to be energy 
independent.
For the  $a_0(980)$, a simple Breit-Wigner formula is used instead of
the more sophisticated formula used in Ref.~\cite{mori1,mori2} for
the $f_0(980)$.
This is because the resonance shape appears to be symmetric with no 
indication of the effect of the $K \bar{K}$ threshold.
In fact, a fit with the formula in Ref.~\cite{mori1, mori2} gives 
$g^2_{K \bar{K}}/g^2_{\eta \pi}=0^{+0.03}_{-0}$,
where $g_{KK}$ ($g_{\pi \pi}$) is the coupling of the $a_0(980)$
to $K \bar{K}$ ($\pi \pi$).

\begin{center}
\begin{table}
\caption{Parameters of the $a_2(1320)$~\cite{pdg}.}
\label{tab:a2para}
\begin{tabular}{ccc} \hline \hline
Parameter & Value & Unit  \\ \hline
 Mass & $1318.3 \pm 0.6$ & $\MeV/c^2$ \\
$\Gamma_{\rm tot}$ & $107 \pm 5$ & MeV \\
${\cal B}({a_2 \rightarrow \rho \pi})$ & $70.1 \pm 2.7$ & \% \\
${\cal B}({a_2 \rightarrow \eta \pi})$ & $14.5 \pm 1.2$ & \% \\
${\cal B}({a_2 \rightarrow \omega \pi \pi})$ & $10.6 \pm 3.2$ & \% \\
${\cal B}({a_2 \rightarrow K \bar{K}})$ & $4.9 \pm 0.8$ & \% \\
${\cal B}({a_2 \rightarrow \gamma \gamma})$ & $(9.4 \pm 0.7) \times 10^{-6}$
& -- \\
\hline \hline
\end{tabular}
\end{table}
\end{center}

\subsubsection{Fitted parameters}
We fit differential cross sections with the
parameterized amplitudes for the range $0.90~\GeV \leq W \leq  1.46~\GeV$.
There are 19 parameters to be fitted.
About one thousand sets of randomly generated initial parameters are prepared 
and fitted using MINUIT~\cite{minuit}
to search for the true minimum and to find any multiple solutions.
A unique solution is found with $\chi^2/ndf = 597.6/429 = 1.39$
($ndf$ denotes the number of degrees of freedom)
for the nominal fit, which appears in more than $\sim 3$\% of the cases.
The fitted parameters are listed in Table~\ref{tab:fit}.
The quoted errors are MINOS statistical errors.
They are calculated from the $\chi^2$ values obtained by varying each 
parameter while floating all the other parameters.
\begin{center}
\begin{table}
\caption{Fitted parameters. The errors are statistical only.} 
\label{tab:fit}
\begin{tabular}{l|lcccc} \hline \hline
Resonance & Parameter  & Nominal  & $M(a_0(Y))$ fixed & No $a_0(Y)$ 
& Unit\\
\hline
$a_0(980)$ & Mass &  $982.3^{+0.6}_{-0.7}$ & $982.3^{+0.8}_{-0.7}$ 
& $982.3 \pm 0.6$ & MeV/$c^2$ \\
& $\Gamma_{\rm tot}$ & $75.6 \pm 1.6$ & $76.9^{+1.0}_{-1.3}$ 
& $75.6^{+1.4}_{-1.3}$ & MeV \\
& $\Gamma_{\gamma \gamma}{\cal B}(\eta \pi^0)$ 
&  $128^{+3}_{-2}$ &  $558^{+52}_{-44}$ & $642 \pm 8$ & eV \\
\hline
$a_0(Y)$ & Mass&  $1316.8^{+0.7}_{-1.0}$ & 1474.0 (fixed) & -- & MeV/$c^2$\\
& $\Gamma_{\rm tot}$ & $65.0^{+2.1}_{-5.4}$ & $251^{+25}_{-33}$ & -- 
& MeV \\
& $\Gamma_{\gamma \gamma} {\cal B}(\eta \pi^0)$ 
&  $432 \pm 6$ &  $(11.0^{+4.4}_{-3.3}) \times 10^3$ & 0 (fixed) 
& eV \\
\hline
\multicolumn{2}{c}{ $\chi^2/ndf$} 
& ~~597.6/429=1.39~~  &  ~~704.5/430=1.65~~  & ~~753.6/433=1.74~~ & \\
\hline\hline
\end{tabular}
\end{table}
\end{center}

Differential cross sections together with the fitted curves are shown 
in Fig.~\ref{fig11}
for selected $W$ bins.
The fit is reasonable as can be seen from these bins
and from Fig.~\ref{fig12}, where
the quantities $\hat{S}^2$ and $\hat{D}_2^2$ are reproduced reasonably well.
\begin{figure}
 \centering
   {\epsfig{file=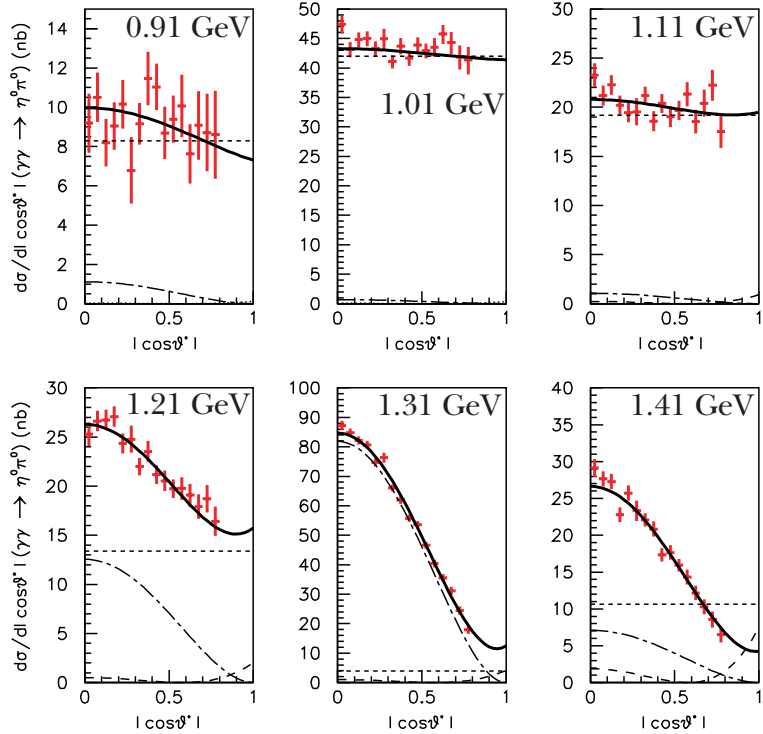,width=100mm}}
 \caption{Differential cross section 
($d \sigma / d |\cos \theta^*|$ (nb)) 
(data points) and results of the fit (solid line) 
for the $W$ bins indicated. 
The dotted, dashed and dot-dashed curves indicate the
$|S|^2$, $4 \pi |D_0 Y_2^0|^2$ and $4 \pi |D_2 Y_2^2|^2$ contributions,
respectively.}
\label{fig11}
\end{figure}

\begin{figure}
 \centering
   {\epsfig{file=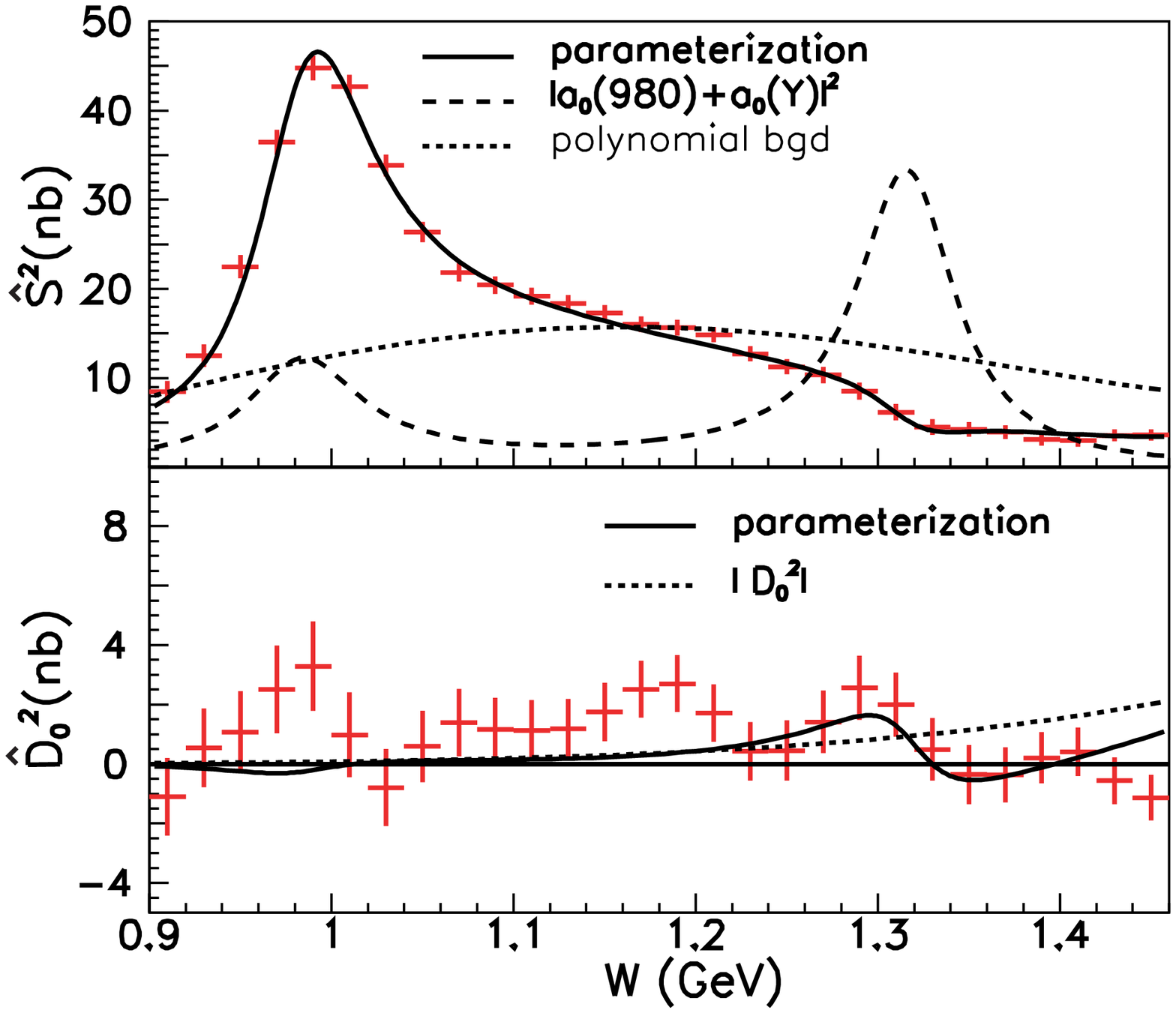,width=60mm}}
   {\epsfig{file=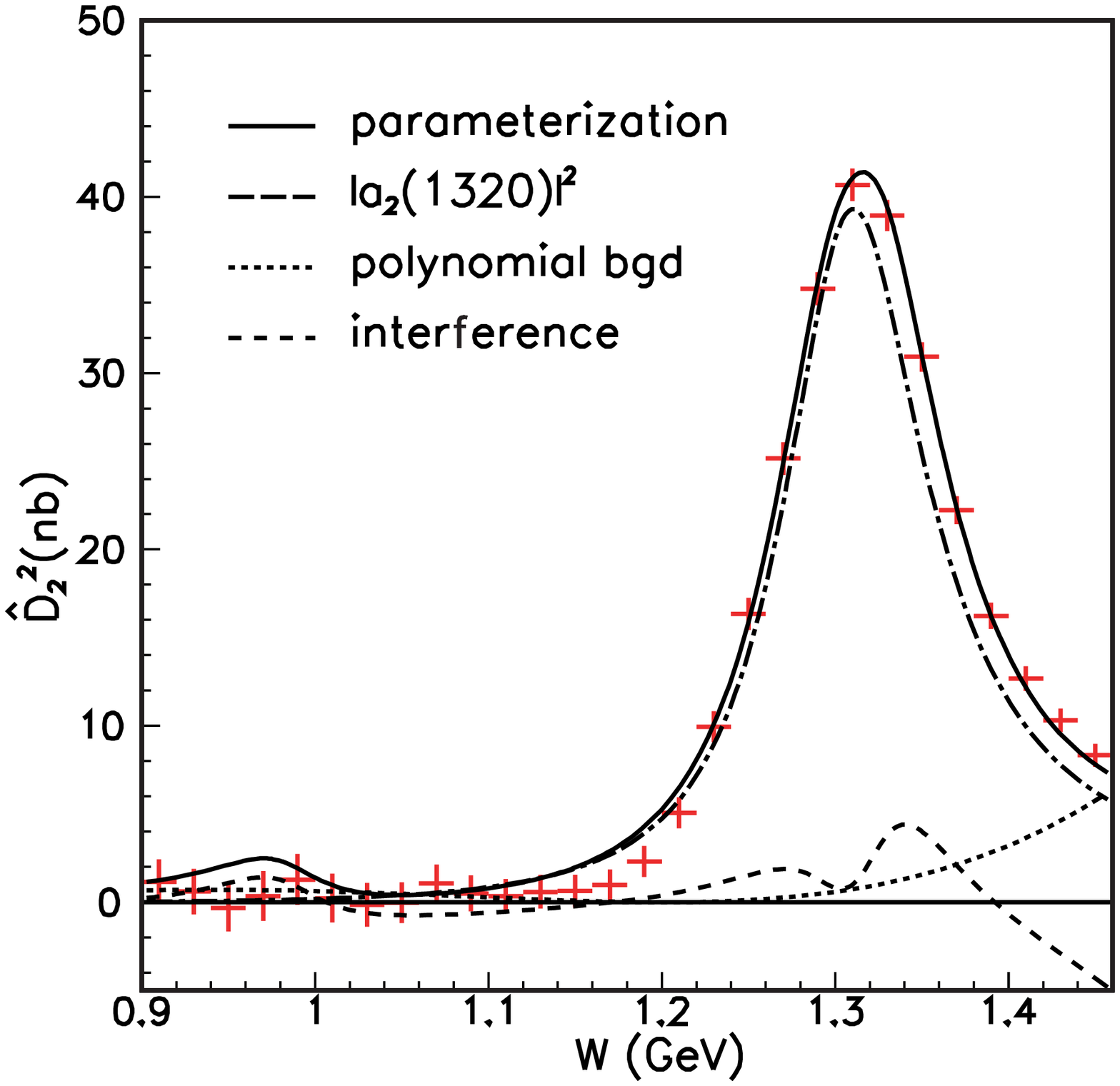,width=60mm}}
 \caption{Results of the parameterization and $\hat{S}^2$ (left top),
$\hat{D}_0^2$ (left bottom), and $\hat{D}_2^2$ (right).
The error bars shown are diagonal statistical error.}
\label{fig12}
\end{figure}

The total cross section ($|\cos \theta^*| < 0.8$) can be obtained
by integrating Eq.~(\ref{eqn:diff2}) as:
\begin{equation}
\sigma_{\rm tot} = 0.8 \hat{S}^2 +  0.457 \hat{D}_0^2
+ 0.983 \hat{D}_2^2 \;,
\end{equation}
where the factors come from the integration of spherical harmonics for
$|\cos \theta^*| \leq 0.8$.
The measured total cross section is in good agreement with the prediction
obtained from the sum of the fitted amplitudes as shown in Fig.~\ref{fig13}.
\begin{figure}
 \centering
   {\epsfig{file=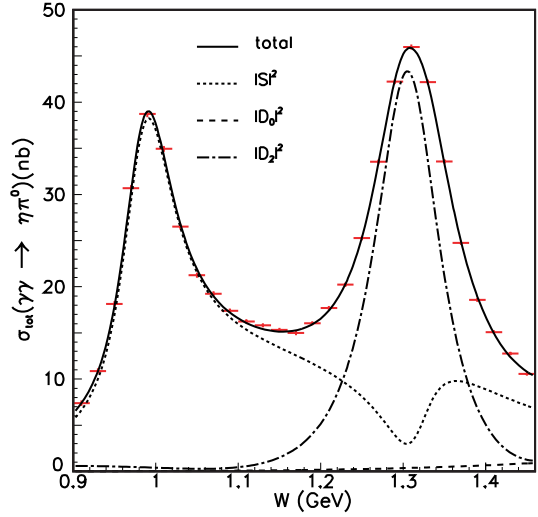,width=70mm}}
 \caption{Total cross section ($|\cos \theta^*| < 0.8$) and 
results of the parameterization.
Contributions of $|S|^2, \; |D_0|^2$ and $|D_2|^2$
(not $\hat{S}^2$, $\hat{D}_0^2$ and $\hat{D}_2^2$) are also shown.}
\label{fig13}
\end{figure}

So far we have used the measured parameter values for the $a_2(1320)$.
However, having high statistics two-photon production data, one might 
question their validity, in particular, 
$\B (a_2(1320) \rightarrow \gamma \gamma)$.
Namely, the determination of this quantity may have been biased
in past experiments by the presence of non-resonant background, etc.
To study this effect, a fit is performed where the product
$\Gamma_{\gamma \gamma} {\cal B}(\eta \pi^0)$ of the 
$a_2(1320)$ is also floated.
The value $\Gamma_{\gamma \gamma}(a_2(1320)) {\cal B}(\eta \pi^0)$ 
obtained is $(145^{+97}_{-34})$~eV, which corresponds to 
$\B (a_2(1320) \rightarrow \gamma \gamma) = (9.4^{+6.3}_{-2.2}) 
\times 10^{-6}$; it agrees well with the PDG value,
$(9.4 \pm 0.7) \times 10^{-6}$ within the rather large fitting error. 
Thus, we conclude that the parameters obtained in past measurements are 
reasonable.
The large statistical error in our measurement arises because of
interference.

The $a_0(Y)$ mass is close to that of the
$a_2(1320)$, which may suggest the possibility that the $a_0(Y)$ is just a 
contribution of the $a_2(1320)$ to the D$_0$ wave, which is not
taken into account in the above fit.
To test this hypothesis, a fit is performed with 
the $a_0(Y)$ removed and with $D_0$ and $D_2$ 
in Eq.~(\ref{eqn:param}) replaced by
\begin{eqnarray}
D_0 &=& \sqrt{\frac{r_{02}}{1 + r_{02}}} A_{a_2(1320)} e^{i \phi_{d0}} 
+ B_{D0} \; , \nonumber \\
D_2 &=& \sqrt{\frac{1}{1 + r_{02}}} A_{a_2(1320)} e^{i \phi_{d2}} 
+ B_{D2} \; ,
\label{eqn:param2}
\end{eqnarray}
where $r_{02}$ indicates the fraction of the $a_2(1320)$ in the
D$_0$ wave.
We obtain $r_{02} = 1.7^{+0.5}_{-0.4}$\% with $\chi^2/ndf = 737.7/431$.
The fit quality is unacceptably poor.
When the $a_0(Y)$ is restored and a fit is performed with 
Eq.~(\ref{eqn:param2}),
we obtain $r_{02} = 3.4^{+2.3}_{-1.1}$\% with $\chi^2/ndf = 580.6/427$.
We conclude that the contribution of the $a_2(1320)$ to the D$_0$ wave
is small.

The mass and width of the $a_0(Y)$ are significantly smaller than
$1474 \pm 19~\MeV/c^2$ and $265 \pm 13~\MeV$, the parameters of 
the $a_0(1450)$ in the PDG~\cite{pdg}.
Since the fit without the $a_0(Y)$ or one where the mass 
of the $a_0(Y)$ fixed to the $a_0(1450)$ mass 
are unacceptable (Table~\ref{tab:fit}),
the inclusion of the $a_0(Y)$ is required
to explain the structure in $\hat{S}^2$ near 
1.3~GeV seen in Fig.~\ref{fig12}.
Since, the mass and width of the $a_0(1450)$ are far from established,
we may either identify the $a_0(Y)$ with the $a_0(1450)$ or
treat it as another scalar resonance.

\subsubsection{Study of systematic errors}
The following sources of systematic errors on the parameters are 
considered: dependence on the fitted region, normalization errors in the 
differential cross sections, assumptions about the background amplitudes,
uncertainties from the unfolding procedure,
and the uncertainties of the $a_2(1320)$.
For each study, a fit is performed allowing all the parameters to float and
the differences of the fitted parameters from the nominal values
are quoted as systematic errors.
A thousand sets of randomly prepared input parameters are prepared
for each study
and fitted to search for the true minimum and for possible multiple solutions.
Unique solutions are found very often.

Two fitting regions are tried: higher
($ 0.94~\GeV \leq W \leq 1.50~\GeV$)
and lower ($ 0.86~\GeV \leq W \leq 1.42~\GeV$).
Normalization error studies are divided into those from uncertainties of 
overall normalization and those from distortion of the spectra in either
$|\cos \theta^*|$ or $W$.
For overall normalization errors, fits are made with two sets of values
of differential cross sections
obtained by multiplying by $(1 \pm \sigma_{\epsilon(W, |\cos \theta^*|)})$, 
where $\sigma_{\epsilon}$ is the relative efficiency error;
the results are denoted ``normalization'' errors.
For distortion studies, $\pm 4$\% errors are assigned and differential 
cross sections are distorted by multiplying them by 
$(1 \pm 0.1 |\cos \theta^*| \mp 0.04)$
and  $(1 \pm (W(\GeV) - 1.18)/7)$
(referred to as ``bias:$|\cos \theta^*|$'' and ``bias:$W$'' errors, 
respectively).

For studies of background (BG) amplitudes, one of the waves 
is changed to a first- or a third-order polynomial.
In addition, constant terms are fixed to non-zero values.
Uncertainties from the unfolding procedure are studied 
by analyzing the differential cross sections where a key parameter 
in unfolding is varied within an allowable range.
Finally, the parameters of the $a_2(1320)$ are successively varied
by their $\pm 1 \sigma$ uncertainties.

The resulting systematic errors are summarized in Table~\ref{tab:syser}.
The total systematic errors are calculated by combining the individual 
errors in quadrature.
As can be seen in Table~\ref{tab:syser},
the values of $\Gamma_{\gamma \gamma} \B (\eta \pi^0)$
for the $a_0(980)$ and $a_0(Y)$ jump to much larger values in some of 
the systematic variations.
This is because destructive interference is preferred in some of the studies.
A study reveals that for much narrower $W$ regions, e.g. about the
$a_0(980)$ peak region, there exist two solutions corresponding to
constructive and destructive interference; 
the latter solution is disfavored in a wider
$W$ range in the nominal fit, but favored in some of the systematic studies.
In this case, taking the sum in quadrature of the individual errors will 
be an overestimation.
Thus, we choose the maximum deviation among the different systematic 
variations to estimate the total systematic error.

{\large
\begin{table}
\caption{Systematic uncertainties for $a_0(980)$ and $a_0(Y)$ parameters}
\label{tab:syser}
\begin{center}
\renewcommand{\arraystretch}{1.2}
\begin{tabular}{l|lll|lll} \hline \hline 
& \multicolumn{3}{c|}{$a_0(980)$} & \multicolumn{3}{c}{$a_0(Y)$} 
\\ \cline{2-7}
Source & Mass & $\Gamma_{\rm tot}$  & $\Gamma_{\gamma \gamma} \B (\eta \pi^0)$ 
 & Mass & $\Gamma_{\rm tot}$ & $\Gamma_{\gamma \gamma} {\cal B}_{\eta \pi0}$\\
& (MeV/$c^2$) & ~~~(MeV)~~~ &~~(eV)~~&(MeV/$c^2$)&(MeV)&(eV) \\ 
\hline
$W$ fit range& $^{+0.4}_{-1.1}$& $^{+2.4}_{-0.0}$& $^{+7.9}_{-0.0}$
& $^{+5.7}_{-1.2}$& $^{+4.3}_{-2.7}$& $^{+0.0}_{-57.0}$\\
Normalization& $^{+0.4}_{-0.6}$& $^{+0.2}_{-0.2}$& $^{+ 17.6}_{-19.5}$
& $^{+0.9}_{-0.0}$& $^{+0.0}_{-1.3}$& $^{+ 20.2}_{-46.4}$\\
Bias:$|~\cos ~\theta^|$& $^{+0.0}_{-0.1}$& $^{+0.4}_{-0.2}$& $^{+0.0}_{-9.3}$
& $^{+3.6}_{-0.0}$& $^{+6.6}_{-18.1}$& $^{+0.0}_{-179.3}$\\
Bias:$W$ & $^{+0.0}_{-0.1}$& $^{+0.2}_{-0.1}$& $^{+5.1}_{-4.6}$
& $^{+0.1}_{-0.0}$& $^{+0.5}_{-0.5}$& $^{+0.8}_{-1.4}$\\
\hline
BG:${\rm Re}(S)$ param. & $^{+2.9}_{-0.5}$& $^{+0.0}_{-1.4}$& $^{+0.0}_{-10.6}$
& $^{+0.0}_{-0.8}$& $^{+ 12.2}_{-0.9}$& $^{+150.2}_{-35.0}$\\
BG:${\rm Im} (S)$ param.& $^{+0.0}_{-1.6}$& $^{+0.5}_{-0.6}$
& $^{+482.4}_{-6.1}$
& $^{+ 17.3}_{-0.0}$& $^{+ 97.8}_{-6.8}$& $^{+ 1061.1}_{-141.5}$\\
BG:${\rm Re}(S) \; c \ne 0$ & $^{+0.1}_{-0.5}$& $^{+0.1}_{-0.3}$
& $^{+0.0}_{-9.3}$
& $^{+0.1}_{-0.1}$& $^{+0.0}_{-0.4}$& $^{+0.0}_{-12.0}$\\
BG:${\rm Re}(D_0)$ param.& $^{+0.0}_{-0.9}$& $^{+0.0}_{-0.3}$& $^{+0.0}_{-2.0}$
& $^{+6.6}_{-0.0}$& $^{+1.9}_{-10.2}$& $^{+ 18.1}_{-215.7}$\\
BG:${\rm Im} (D_0)$ param.& $^{+0.0}_{-0.9}$& $^{+0.0}_{-0.3}$
& $^{+0.0}_{-2.0}$
& $^{+6.6}_{-0.0}$& $^{+0.0}_{-10.3}$& $^{+0.0}_{-215.7}$\\
BG:${\rm Re}(D_0) \; c \ne 0$ 
& $^{+0.3}_{-0.7}$& $^{+0.3}_{-0.0}$& $^{+0.0}_{-9.4}$
& $^{+6.6}_{-0.0}$& $^{+0.3}_{-14.4}$& $^{+0.0}_{-229.6}$\\
BG:${\rm Re}(D_2)$ param.& $^{+0.0}_{-0.1}$& $^{+0.5}_{-0.0}$& $^{+0.0}_{-6.3}$
& $^{+5.4}_{-0.0}$& $^{+0.0}_{-13.4}$& $^{+0.0}_{-210.8}$\\
BG:${\rm Im} (D_2)$ param.& $^{+0.0}_{-0.0}$& $^{+0.0}_{-0.0}$
& $^{+0.2}_{-3.0}$
& $^{+0.6}_{-0.0}$& $^{+0.0}_{-1.3}$& $^{+0.0}_{-18.4}$\\
BG:${\rm Re}(D_2) \; c \ne 0$ 
& $^{+0.5}_{-0.0}$& $^{+0.5}_{-0.0}$& $^{+3.9}_{-0.6}$
& $^{+0.6}_{-0.0}$& $^{+1.2}_{-0.3}$& $^{+ 11.7}_{-6.6}$\\
\hline
Unfolding & $^{+0.7}_{-3.9}$& $^{+ 17.2}_{-9.8}$& $^{+501.2}_{-31.3}$
& $^{+9.7}_{-3.2}$& $^{+0.0}_{-7.4}$& $^{+0.0}_{-213.3}$\\
\hline
$a_2$:mass & $^{+0.0}_{-0.1}$& $^{+0.1}_{-0.1}$& $^{+0.7}_{-0.8}$
& $^{+0.9}_{-0.9}$& $^{+0.0}_{-0.3}$& $^{+0.7}_{-5.6}$\\
$a_2$:width& $^{+0.2}_{-0.2}$& $^{+1.0}_{-0.8}$& $^{+3.9}_{-4.1}$
& $^{+2.7}_{-2.8}$& $^{+5.4}_{-4.7}$& $^{+ 48.0}_{-38.9}$\\
$a_2: \Gamma_{\gamma \gamma}{\cal B}(\eta \pi^0)$ 
& $^{+0.0}_{-0.1}$& $^{+0.1}_{-0.0}$& $^{+0.7}_{-0.6}$
& $^{+1.1}_{-0.6}$& $^{+0.7}_{-1.6}$& $^{+ 12.2}_{-25.3}$\\
$a_2: r_R$& $^{+0.3}_{-0.0}$& $^{+0.4}_{-0.0}$& $^{+5.3}_{-0.0}$
& $^{+0.4}_{-0.3}$& $^{+2.6}_{-1.6}$& $^{+ 8.4}_{-10.9}$\\
\hline
Total& $^{+3.1}_{-4.7}$& $^{+ 17.4}_{-10.0}$& $^{+501.6}_{-43.0}$
& $^{+ 24.7}_{-4.6}$& $^{+ 99.1}_{-32.6}$& $^{+ 1073.2}_{-255.5}$\\
\hline  \hline 
\end{tabular}
\end{center}
\end{table}
}

\subsubsection{Summary of resonance studies}
To summarize the study presented in this section,
once the amplitudes are parameterized, differential cross sections can be
fitted to obtain the parameters as described above.
Although the fit itself is not very good as can be seen from 
$\chi^2/ndf = 1.39$, it is stable despite the fact that the approach to
the minimum is slow; tens of MINUIT runs are needed to reach the minimum.
The mass, width and $\Gamma_{\gamma \gamma} \B (\eta \pi^0)$ values obtained 
for the $a_0(980)$ and $a_0(Y)$ 
are summarized and compared to those in the PDG~\cite{pdg}
in Tables~\ref{tab:a098} and \ref{tab:a0y}.
Note that the value of the product
$\Gamma_{\gamma \gamma} \B (\eta \pi^0)$ in the
PDG~\cite{pdg} is an average of Refs.~\cite{cbep} and \cite{jade}.
In both analyses, the total cross section or an event distribution is
fitted to an incoherent sum of the $a_0(980)$ and $a_2(1320)$
resonances (with the masses and widths fixed to the earlier PDG 
values~\cite{oldpdg})
and nonresonant background because of the limited statistics available.
In our fit, we fully take into account interference among amplitudes.
When we follow the same procedure as in the previous analyses, 
which ignored possible interference, we reproduce 
their values with much better statistical errors.

\begin{center}
\begin{table}
\caption{Fitted parameters of the $a_0(980)$}
\label{tab:a098}
\begin{tabular}{cccc} \hline \hline
Parameter & This work & PDG & Unit  \\ \hline
 Mass & ~~~~$982.3~^{+0.6}_{-0.7}~^{+3.1}_{-4.7}$~~~~
& ~~~~$984.7 \pm 1.2$~~~~
& $\MeV/c^2$ \\
$\Gamma_{\rm tot}$ & $75.6~\pm 1.6~^{+17.4}_{-10.0}$ & 50 -- 100 & MeV \\
$\Gamma_{\gamma \gamma} \B (\eta \pi^0)$ &
$128~^{+3}_{-2}~^{+502}_{-43}$ & $240~^{+80}_{-70}$ & eV\\
\hline \hline
\end{tabular}
\end{table}
\end{center}

\begin{center}
\begin{table}
\caption{Fitted parameters of the $a_0(Y)$ compared to those of the
 $a_0(1450)$}
\label{tab:a0y}
\begin{tabular}{cccc} \hline \hline
Parameter & This work & $a_0(1450)$ (PDG) & Unit  \\ \hline
 Mass & ~~~~$1316.8~^{+0.7}_{-1.0}~^{+24.7}_{-4.6}$~~~~
& ~~~~$1474 \pm 19$ $\ \ $ & $\MeV/c^2$ \\
$\Gamma_{\rm tot}$ & $65.0~^{+2.1}_{-5.4}~^{+99.1}_{-32.6}$ & 
$265 \pm 13$ & MeV \\
$\Gamma_{\gamma \gamma} \B (\eta \pi^0)$ &
$432~\pm 6~^{+1073}_{-256}$ & unknown & eV\\
\hline \hline
\end{tabular}
\end{table}
\end{center}

\section{Analysis of the higher energy region}
\label{sec-5}
In this section, we study the angular dependence of the differential cross 
sections, $W$ dependence of the total cross section,
and the ratio of cross sections for $\eta \pi^0$ to $\pi^0 \pi^0$
production in the high energy region, $W > 2.4~\GeV$.

\subsection{Angular dependence}
As in the analysis of the $\pi^0\pi^0$ process~\cite{pi0pi02}, 
we compare the angular dependence of the differential cross sections 
with the function $\sin^{-4}{\theta^*}$. 
A fit with an additional $\cos^2{\theta^*}$ term does not significantly
improve the fit quality. 
Limited statistics prevent us from quantifying a possible deviation
from the  $\sin^{-4}{\theta^*}$ behavior when we study the $W$ dependence of
the data.
Here, we only show comparison with a
$\sin^{-4} \theta^*$ parameterization in different $W$ regions in 
Fig.~\ref{fig14}.
In this figure, the vertical axis is the differential cross section 
divided by the total integral over
$|\cos \theta^*|<0.8$.
The curve is $0.602 \sin^{-4} \theta^*$ (not a fit).
The numerical factor is the differential cross section 
normalized to $\sigma (|\cos \theta^*|<0.8)$.
The experimental result shows that the agreement is good for $W > 2.7~\GeV$.

\subsection{Power-law $W^{-n}$ dependence}
We fit the $W$ dependence of the total cross section ($|\cos \theta^*|<0.8$) 
in the energy region 3.1-4.0~GeV, where the lower boundary 3.1~GeV is 
the same as in the $\pi^0\pi^0$ analysis. 
The fit gives $n=10.5 \pm 1.2 \pm 0.5$,
and the corresponding cross section is drawn in Fig.~\ref{fig15}(a) as well as
that of the $\pi^0\pi^0$ process in the same angular range.  
The systematic error is 
obtained from the difference of the central values when we shift the
cross section by $\pm 1 \sigma$ at 3.1~GeV and $\mp 1 \sigma$ at 4.0~GeV
and by factors obtained by connecting linearly for $W$ bins in between,
where $\sigma$ is an energy-dependent part of systematic error at 
each $W$ point.
The $n$ value can be compared with $n$ values in other processes that we
studied earlier~\cite{nkzw,wtchen,pi0pi02}.
The results are summarized in Table~\ref{tab:val_n} from which
it is clear that the result for the $\eta\pi^0$ final 
state is consistent with that for $K^0_S K^0_S$ (where the fitted $W$ range is 
wider), but two standard deviations higher than that for $\pi^0\pi^0$. 
The energy dependence of the latter 
seems to be different from that of the two other purely neutral final 
states and closer to $\pi^+\pi^-$ and $K^+K^-$, although no strict 
conclusions can be drawn at this level of statistics.

\begin{center}
\begin{table}
\caption{The value of $n$ in $\sigma_{\rm tot} \propto W^{-n}$ in
various reactions fitted in the $W$ and $|\cos \theta^*|$ ranges indicated.}
\label{tab:val_n}
\begin{tabular}{lcccc} \hline \hline
Process & $n$ & $W$ range (GeV) & $|\cos \theta^*|$ range & Reference
\\ \hline
$\eta \pi^0$ & $10.5 \pm 1.2 \pm 0.5$ & 3.1 -- 4.1 & $<0.8$ & This work \\
$\pi^0\pi^0$ & $8.0 \pm 0.5 \pm 0.4$ & $ \ \ \ 
$ 3.1 -- 4.1 (exclude 3.3 -- 3.6)
$ \ \ \ $  & $<0.8$ & \cite{pi0pi02} \\
$K^0_S K^0_S$  & $10.5 \pm 0.6 \pm 0.5$ & 2.4 -- 4.0 (exclude 3.3 -- 3.6) 
& $<0.6$ & \cite{wtchen} \\
\hline
$\pi^+\pi^-$ & $7.9 \pm 0.4 \pm 1.5$ & 3.0 -- 4.1 & $<0.6$ & \cite{nkzw} \\
$K^+K^-$  & $7.3 \pm 0.3 \pm 1.5$ & 3.0 -- 4.1 & $<0.6$ & \cite{nkzw} \\
\hline\hline
\end{tabular}
\end{table}
\end{center}

\subsection{Cross section ratio}
A ratio of cross sections among  neutral-pseudoscalar-meson 
($\pi^0$ or $\eta$) pair production in two-photon collisions
can be predicted relatively easily within a pQCD model.
The pQCD model in Ref.~\cite{bl} predicts the cross section ratio 
$\sigma(\eta\pi^0) / \sigma(\pi^0\pi^0)$
as summarized in Table~\ref{tab:ratpre}. 
In the table, $R_f = (f_{\eta}/f_{\pi^0})^2$, where $f_\eta$ ($f_\pi$)
is the $\eta$ ($\pi^0$) form factor;
the value of $R_f$ is not well known and we temporarily  assume it to be unity.
The ratio of the cross sections is proportional to the
square of the coherent sum of the product of the quark charges, 
$|\Sigma e_1 e_2|^2$, in which $e_1 = -e_2$ in the
present neutral-meson production cases. 
We show two predictions: a pure flavor SU(3) octet state 
and a mixture with $V_P=-18^\circ$ for the $\eta$ meson.
For comparison,
we also show in the table the values calculated using an incoherent sum
as an example of an extreme case.
Here, we assume that the quark-quark component of the neutral meson wave 
functions dominates and is much larger than
the two-gluon component,
in obtaining the relations between the cross sections.

\begin{center}
\begin{table}
\caption{Predictions for the cross section ratio:
$\sigma(\eta\pi^0) / \sigma(\pi^0\pi^0)$
in two-photon collisions. 
``Coherent sum'' (``Incoherent sum'') is the ratio
derived from the squared coherent (incoherent) sum of the product of
the constituent-quark charges. 
Here, $R_f = (f_{\eta}/f_{\pi^0})^2$, where $f_\eta$ ($f_\pi$)
is the $\eta$ ($\pi^0$) form factor; the value may be taken to be $R_f = 1$.
The $\eta$ meson is treated as a pure SU(3) octet state for the 
entries in the ``octet'' row,
while ``$V_P=-18^\circ$'' is the most probable mixing angle between
the octet and singlet states from experiments.\\}
\label{tab:ratpre}
\begin{tabular}{c|c|c}
\hline \hline
$\eta$ in SU(3) & Coherent sum 
& Incoherent sum\\ 
\hline
Octet 
& ~~~$0.24 R_f$~~~ 
& ~~~$0.67 R_f$~~~ 
\\
$V_P=-18^\circ$ 
& $0.46 R_f$ 
& $1.29 R_f$ 
\\      \hline \hline
\end{tabular}
\end{table}
\end{center}

The $W$ dependence of the ratio between the measured cross section 
integrated over $|\cos \theta^*|<0.8$ of 
$\gamma \gamma \to \eta \pi^0$ to
$\gamma \gamma \to \pi^0 \pi^0$ is plotted in Fig.~\ref{fig15}(b).
For the $\pi^0\pi^0$ process, the
contributions from charmonium production are subtracted using a 
model-dependent assumption described in Ref.~\cite{pi0pi02}.
Even though the ratio may have a slight $W$ dependence,
we average the ratio of the
cross sections over the range $3.1~\GeV < W < 4.0~\GeV$ as was done in other
processes for the sake of comparison with QCD and 
obtain $0.48 \pm 0.05 \pm 0.04$. 
In the averaging, the ratio in the charmonium region (in 
$\pi^0\pi^0$) 3.3 - 3.6~GeV is not used. 
This ratio is in agreement with the QCD prediction 
if we take $R_f = 1$.

\subsection{Comments on charmonium}
  It is conjectured that the known $c\bar{c}$ charmonium 
states do not decay into
the $\eta\pi^0$ final state with any observable rate, because the 
$I=1$ component should be suppressed. 
In other words, this process could be useful to 
search for a new charmonium-like particle  with $I=1$ that would be a 
candidate for an exotic resonance.
In Fig.~\ref{fig16}, we show the invariant mass distribution of 
$\eta\pi^0$ events with $|\cos \theta^*|<0.4$,
where a resonance contribution would be enhanced.
We do not observe any signals of the known $\chi_{cJ}$ mesons.  
There is a hint of a peak near 3.18~GeV,
however, its statistical significance is less than $3\sigma$.
Therefore, we assumed in the above
discussion that there are no charmonium contributions in the
measurements of the cross section described here.

\begin{figure}
 \centering
   {\epsfig{file=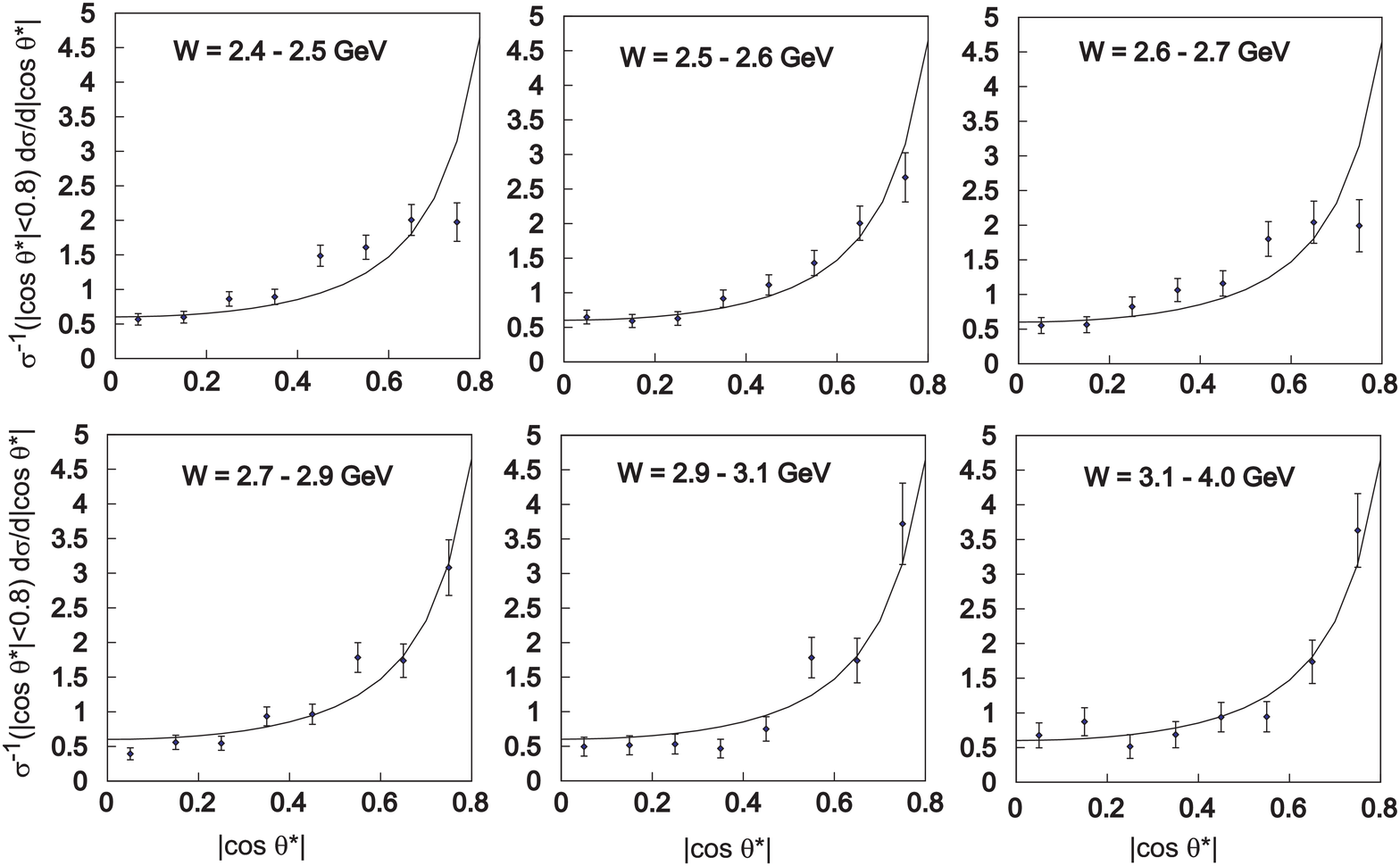,width=110mm}}
 \caption{The angular dependence of the differential cross sections in 
different $W$ regions, with
the normalization to the cross section integrated over $|\cos \theta^*|<0.8$.
The curves are proportional to $\sin^{-4} \theta^*$ and normalized
similarly.}
\label{fig14}
\end{figure}

\begin{figure}
 \centering
   {\epsfig{file=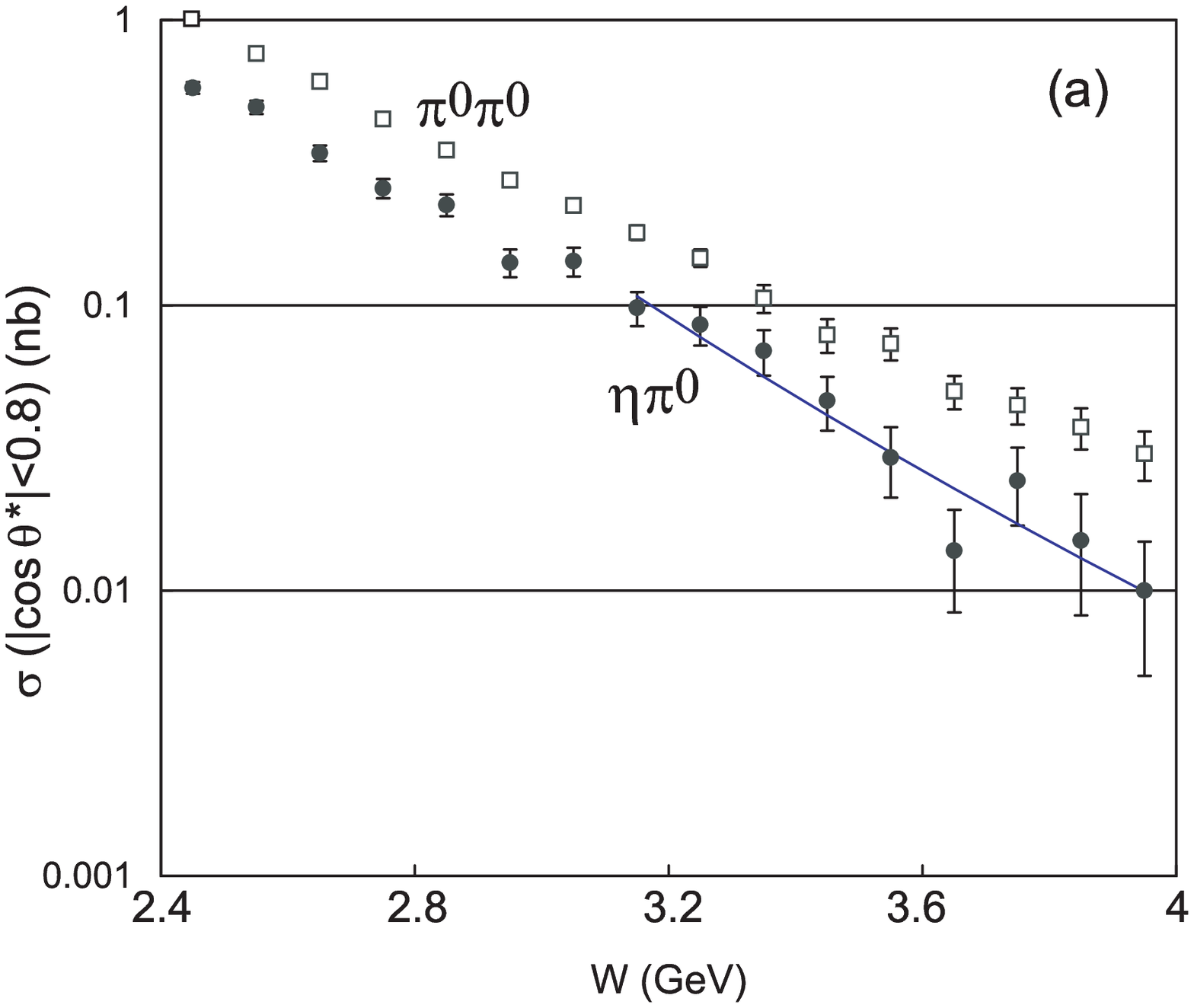,width=60mm}}
   {\epsfig{file=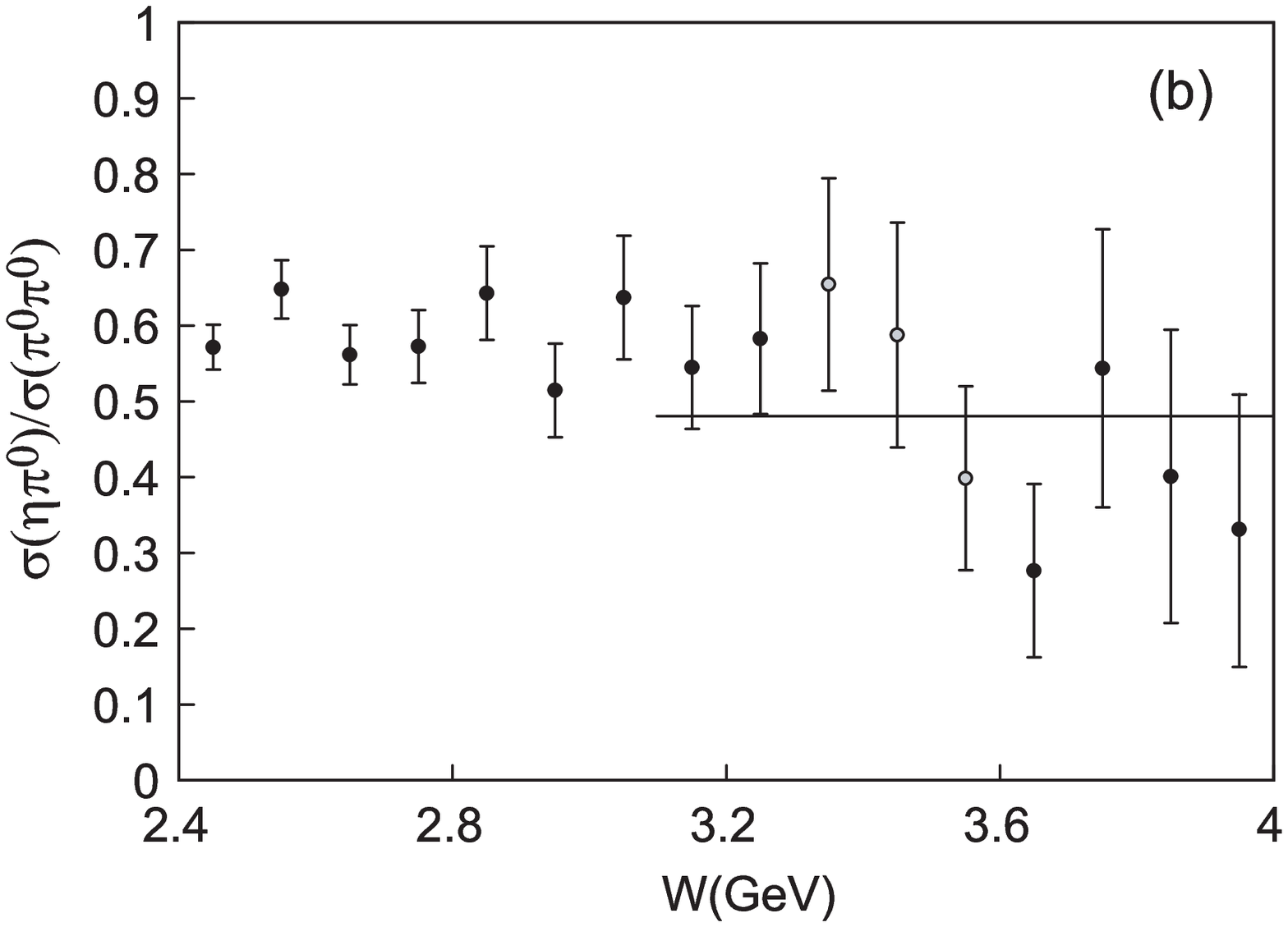,width=60mm}}
 \caption{The $W$ dependence of the cross section ($|\cos \theta^*|<0.8$)
(a). 
The curve is the power-law fit. 
The cross section of $\eta\pi^0$ production is compared to
that for $\pi^0\pi^0$~\cite{pi0pi02}.
The $W$ dependence of the cross section ratio of $\eta\pi^0$ 
to $\pi^0\pi^0$ ($|\cos \theta^*|<0.8$) (b).
The line is the average in the 3.1 - 4.0~GeV range (the charmonium region,
3.3 - 3.6~GeV, is omitted from the calculation~\cite{pi0pi02}).}
\label{fig15}
\end{figure}

\begin{figure}
 \centering
   {\epsfig{file=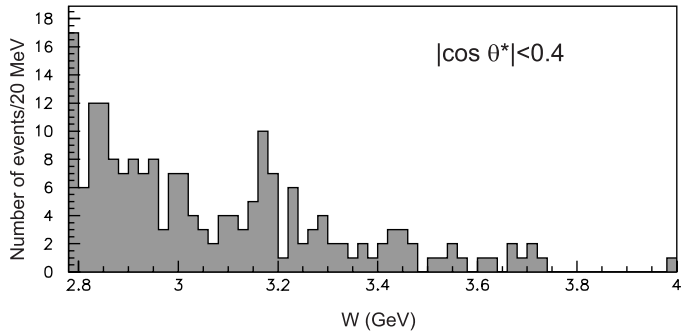,width=90mm}}
 \caption{The $W$ distribution in the charmonium region
($|\cos \theta^*|<0.4$).}
\label{fig16}
\end{figure}

\section{Summary and Conclusion}
\label{sec-6}
We have measured the process $\gamma \gamma \to \eta\pi^0$
using a high-statistics data sample
from $e^+e^-$ collisions corresponding to an integrated luminosity
of 223~fb$^{-1}$ with the Belle detector at the KEKB accelerator. 
We obtain results for the differential cross sections
in the center-of-mass energy and polar angle ranges, 
$0.84~\GeV < W < 4.0~\GeV$ and $|\cos \theta^*|<0.8$. 

Differential cross sections are fitted in the energy region 
$0.90~\GeV \leq W \leq 1.46~\GeV$ in a model where partial
waves consist of resonances and smooth backgrounds.
The D$_0$ wave is small, the D$_2$ wave is dominated by
the $a_2(1320)$ resonance, the S-wave prefers to have at least one
additional resonance (denoted as $a_0(Y)$)
in addition to the $a_0(980)$.
The mass, width and the product $\Gamma_{\gamma \gamma} \B (\eta \pi^0)$
for the $a_0(980) $ are fitted to be 
$982.3~^{+0.6}_{-0.7}~^{+3.1}_{-4.7}~\MeV/c^2$,
$75.6~\pm 1.6~^{+17.4}_{-10.0}~\MeV$,
$128~^{+3}_{-2}~^{+502}_{-43}~\eV$ and
those for the $a_0(Y)$ are
$1316.8~^{+0.7}_{-1.0}~^{+24.7}_{-4.6}~\MeV/c^2$,
$65.0~^{+2.1}_{-5.4}~^{+99.1}_{-32.6}~\MeV$,
$432~\pm 6~^{+1073}_{-256}~\eV$.
The large systematic errors, in particular for two-photon widths, originate 
from an additional solution
that favors destructive interference in some of systematic studies.
The mass and width of the $a_0(Y)$ are significantly smaller than those of the
$a_0(1450)$.
The fact that the obtained $a_0(Y)$ mass is close to the $a_2(1320)$ mass
may suggest that the $a_2(1320)$ contribution in the D$_0$ wave is
important.
However, a fit reveals that the fraction of the $a_2(1320)$ 
in the D$_0$ wave is small and it cannot
replace the $a_0(Y)$.
Since the fit without it or the fit where its mass is fixed 
to the $a_0(1450)$ mass is unacceptable, it is at least a good empirical 
parameterization.
We may still identify the $a_0(Y)$ with the $a_0(1450)$, given
that the latter is far from established, or as another new
scalar meson.
We cannot draw a definite conclusion on the existence of the $a_2(1700)$.

The angular distribution of the differential cross sections is 
close to $\sim \sin^{-4} \theta^*$ above $W=3.1$~GeV
similarly to the $\pi^0\pi^0$.
In this energy region, the energy dependence of the cross section 
integrated over $|\cos \theta^*|<0.8$ is well fitted by
$W^{-n}$, $n=10.5 \pm 1.5 \pm 0.4$,
somewhat higher (by two standard deviations) than that in the
$\pi^0\pi^0$ channel. 
Although a slight $W$ dependence may remain in the ratio, 
we average the cross section ratio, 
$\sigma(\eta \pi^0)/\sigma(\pi^0 \pi^0)$ in the range
$3.1~\GeV < W < 4.0~\GeV$
and obtain $0.48 \pm 0.05 \pm 0.04$.
This ratio is consistent with the prediction from a QCD model based on
$q\bar{q}$ production and SU(3) symmetry. 

\section*{Acknowledgments}
%
We are grateful to V.~Chernyak, M.~Diehl and P.~Kroll for useful discussions.
We thank the KEKB group for the excellent operation of the
accelerator, the KEK cryogenics group for the efficient
operation of the solenoid, and the KEK computer group and
the National Institute of Informatics for valuable computing
and SINET3 network support.  We acknowledge support from
the Ministry of Education, Culture, Sports, Science, and
Technology (MEXT) of Japan, the Japan Society for the 
Promotion of Science (JSPS), and the Tau-Lepton Physics 
Research Center of Nagoya University; 
the Australian Research Council and the Australian 
Department of Industry, Innovation, Science and Research;
the National Natural Science Foundation of China under
contract No.~10575109, 10775142, 10875115 and 10825524; 
the Department of Science and Technology of India; 
the BK21 program of the Ministry of Education of Korea, 
the CHEP src program and Basic Research program (grant 
No. R01-2008-000-10477-0) of the 
Korea Science and Engineering Foundation;
the Polish Ministry of Science and Higher Education;
the Ministry of Education and Science of the Russian
Federation and the Russian Federal Agency for Atomic Energy;
the Slovenian Research Agency;  the Swiss
National Science Foundation; the National Science Council
and the Ministry of Education of Taiwan; and the U.S.\
Department of Energy.
This work is supported by a Grant-in-Aid from MEXT for 
Science Research in a Priority Area ("New Development of 
Flavor Physics"), and from JSPS for Creative Scientific 
Research ("Evolution of Tau-lepton Physics").

\end{document}